\documentclass{ws-ijmpd}
\usepackage[super,compress]{cite}
\usepackage{color,ulem}
\begin{document}

\markboth{Anna Nakonieczna, {\L}ukasz Nakonieczny and Dong-han Yeom}
{Black hole factory: a review of double-null formalism}

%
\catchline{}{}{}{}{}
%

\title{Black hole factory: a review of double-null formalism}

\author{Anna Nakonieczna}

\address{
Institute of Theoretical Physics, Faculty of Physics, University of Warsaw\\
ul. Pasteura 5, 02-093 Warszawa, Poland\\
Anna.Nakonieczna@fuw.edu.pl}

\author{{\L}ukasz Nakonieczny}

\address{
Institute of Theoretical Physics, Faculty of Physics, University of Warsaw\\
ul. Pasteura 5, 02-093 Warszawa, Poland\\
Lukasz.Nakonieczny@fuw.edu.pl}

\author{Dong-han Yeom}

\address{Asia Pacific Center for Theoretical Physics\\
67 Cheongam-ro, Nam-gu, Pohang 37673, Republic of Korea\\
Department of Physics, POSTECH\\
67 Cheongam-ro, Nam-gu, Pohang 37673, Republic of Korea\\
innocent.yeom@gmail.com}

\maketitle


\begin{abstract}
In this review paper, we comprehensively summarize numerical applications of double-null formalism for studying dynamics within the theory of gravity. By using the double-null coordinates, we can investigate dynamical black holes and gravitational phenomena within spherical symmetry, including gravitational collapse, formation of horizons and singularities, as well as evaporations. This formalism can be extended to generic situations, where we can change dimensions, topologies, the gravity sector, as well as the matter sector. We also discuss its possible implications for black hole physics and particle astrophysics. This strong numerical tool will have lots of future applications for various research areas including general relativity, string theory, and various approaches to quantum gravity.
\end{abstract}

%

\newpage

\tableofcontents

\section{Introduction}

General relativity is a very successful theory of gravitation from astrophysical scales to cosmology. However, even with a great success of the theory, we know that there are still lots of limitations involved. Some of such problems can be summarized as follows:
\begin{itemize}
\item[--] \textit{Singularity}: General relativity predicts an existence of a singularity, e.g., inside a black hole and in the early universe \cite{Hawking:1969sw}. This requires that gravity should be either modified or quantized in the strong field regime.
\item[--] \textit{Information loss}: If we include quantum effects of a curved spacetime, one can see evaporation of a black hole \cite{Hawking:1974sw}. Then, after the black hole totally evaporates, will the entire process be unitary \cite{Hawking:1976ra}? This has generated lots of controversial issues for several decades \cite{Chen:2014jwq}.
\item[--] \textit{Dark energy and dark matter}: Observationally, now we know that our universe is dominated by dark matter and dark energy, while we do not know the exact origins of them. Then, what are the origins and how can we confirm their properties? 
\end{itemize}

Considering the above aspects, we need to extend our gravity theory beyond general relativity. However, it is fair to say that we should not seriously bias from general relativity even if we investigate quantum gravity or modified gravity. In terms of dark matter and dark energy, perhaps a promising way to probe them is to use gravitation, and this would require a strong gravitational phenomenon. So, all of the frontiers of gravity are about the strong and dynamical gravitational phenomenon, where exact calculations of higher curvature region will give a guideline toward quantum gravity or beyond general relativity.

The exact calculations of highly dynamical behaviors of gravity are very difficult since one needs to solve time-dependent non-linear equations. Even worse is that the genuine quantum gravitational effect will happen inside the event horizon. Then the accuracy of numerical computations becomes very important. There have been several attempts to study fully dynamical black holes for limited systems \cite{Callan:1992rs}, but some of them are far from realistic four dimensional general relativity.

In this context, the double-null formalism is a good guideline toward dynamical and strong gravity. A comprehensive review of this fascinating tool is the purpose of this paper. We will discuss it later in detail, but one can first summarize the strong points of the double-null formalism as follows:
\begin{itemize}
\item[--] \textit{Free from the coordinate singularity}: One does not need to reparametrize coordinates in order to avoid the coordinate singularity.
\item[--] \textit{Simple boundary conditions}: Boundary is null and hence there is a causal dependence. This makes a simple logical way to give a consistent boundary condition.
\item[--] \textit{Applicability}: Within the same framework, one can easily introduce various extensions.
\end{itemize}

This paper is organized as follows. In Sec. \ref{sec:the}, we first explicitly show the simplest model of the Einstein gravity with a real scalar field. In Sec. \ref{sec:ext}, we discuss extensions of the double-null formalism, at least for four aspects: modifying matter and gravity sectors, dimensions, and topologies. In Sec. \ref{sec:app}, we discuss possible applications of the double-null numerical simulations for black hole physics, cosmology, and particle astrophysics. Finally, in Sec. \ref{sec:per}, we summarize possible future applications and perspectives. In this paper, we use the convention: $c = \hbar = G = 1$.

\section{\label{sec:the}The simplest model}

In this section, we summarize the simplest model: general relativity with a real scalar field \cite{cqg13(1996)497}. This shows clearly how the double-null formalism works both analytically and numerically.

\subsection{\label{sec:fieldeqns}Field equations}

We begin with the Einstein gravity with a scalar field:
\begin{eqnarray}\label{eqn:the-lag}
S = \int dx^{4} \sqrt{-g} \left[ \frac{1}{16\pi} R - \frac{1}{2} g^{\mu\nu} \phi_{;\mu} \phi_{;\nu} \right],
\end{eqnarray}
where $R$ is the Ricci scalar and $\phi$ is a scalar field. Then the equations of motion are the Einstein equation and the Klein-Gordon equation:
\begin{eqnarray}
G_{\mu\nu} = 8 \pi T_{\mu\nu}\;\;\;\;\; \mathrm{and} \;\;\;\;\; \phi_{;\mu\nu}g^{\mu\nu} = 0,
\end{eqnarray}
respectively, where the semicolon denotes the covariant derivative and the stress-energy tensor is
\begin{eqnarray}
T_{\mu\nu} = \phi_{;\mu}\phi_{;\nu} - \frac{1}{2} \phi_{;\rho} \phi_{;\sigma} g^{\rho\sigma} g_{\mu \nu}.
\end{eqnarray}

In the double-null formalism, we use the most generic metric ansatz with the spherical symmetry\footnote{For discussions beyond the spherical symmetry, see \cite{Lehner:1999xi}.}:
\begin{eqnarray}
ds^{2} = -\alpha^{2}(u,v) du dv + r^{2}(u,v) d\Omega^{2},
\end{eqnarray}
where we use the coordinates $[u, v, \theta, \varphi]$, and $u$ and $v$ denote ingoing and outgoing null directions, respectively. From now, for convenience, we denote $\alpha_{,u}/\alpha \equiv h$, $\alpha_{,v}/\alpha \equiv d$, $r_{,u} \equiv f$, $r_{,v} \equiv g$, $\sqrt{4\pi} \phi \equiv s$, $s_{,u} \equiv w$, and $s_{,v} \equiv z$. Then, the components of the Einstein tensor $G_{\mu\nu}$ are as follows:
\begin{eqnarray}
&& G_{uu} = -\frac{2}{r} \left(f_{,u}-2fh \right), \;\;\;\;\; G_{uv} = \frac{1}{2r^{2}} \left( 4 rf_{,v} + \alpha^{2} + 4fg \right),\\
&& G_{vv} = -\frac{2}{r} \left(g_{,v}-2gd \right), \;\;\;\;\; G_{\theta\theta} = -4\frac{r^{2}}{\alpha^{2}} \left(d_{,u}+\frac{f_{,v}}{r}\right),
\end{eqnarray}
while the components of the energy-momentum tensor $T_{\mu\nu}$ are:
\begin{eqnarray}
T_{uu} = \frac{1}{4\pi} w^{2},\;\;\;\;\; T_{uv} = 0,\;\;\;\;\; T_{vv} = \frac{1}{4\pi} z^{2},\;\;\;\;\; T_{\theta\theta} = \frac{r^{2}}{2\pi\alpha^{2}} wz.
\end{eqnarray}
Note that due to the spherical symmetry, the $\theta\theta$- and $\varphi\varphi$-components are proportional each other, $G_{\theta\theta} = \sin^{-2}\theta G_{\varphi\varphi}$, and hence we do not need to regard the $\varphi\varphi$-component as an independent degree of freedom.

Finally, we obtain the explicit form of the field equations, the Einstein equations
\begin{eqnarray}
\label{eq:E1}f_{,u} &=& 2 f h - 4 \pi r T_{uu},\\
\label{eq:E2}g_{,v} &=& 2 g d - 4 \pi r T_{vv},\\
\label{eq:E3}f_{,v}=g_{,u} &=& -\frac{\alpha^{2}}{4r} - \frac{fg}{r} + 4\pi r T_{uv},\\
\label{eq:E4}h_{,v}=d_{,u} &=& -\frac{2\pi \alpha^{2}}{r^{2}}T_{\theta\theta} - \frac{f_{,v}}{r},
\end{eqnarray}
and the Klein-Gordon equation
\begin{eqnarray}
\label{eq:s}z_{,u} = w_{,v} &=& - \frac{fz}{r} - \frac{gw}{r}.
\end{eqnarray}

Now, we comment on a number of variables and equations. We have three variables that we need to solve for: $r$, $\alpha$, and $\phi$. If we have equations for $r_{,uv}$ (Eq. (\ref{eq:E3})), $\alpha_{,uv}$ (Eq. (\ref{eq:E4})), and $\phi_{,uv}$ (Eq. (\ref{eq:s})), then we can decide the variables up to boundary conditions. However, we have two more equations for $r_{,uu}$ (Eq. (\ref{eq:E1})) and $r_{,vv}$ (Eq. (\ref{eq:E2})). These constraint equations will be used to assign consistent boundary conditions. Of course, it is possible to use Eqs.~(\ref{eq:E1}) and ~(\ref{eq:E2}) not only for constraints but also for evolutions.

\subsection{Boundary conditions}

As all the equations are first order differential equations, we need initial conditions for functions $\alpha, h, d, r, f, g, s, w, z$ on the initial $u=u_{\mathrm{i}}$ and $v=v_{\mathrm{i}}$ surfaces, where we set $u_{\mathrm{i}}=v_{\mathrm{i}}=0$ without loss of generality.

First, we have gauge freedom to choose $r$. Here, we choose $r(0,0)=r_{0}$, $f(u,0)=r_{u0}$, and $g(0,v)=r_{v0}$, where $r_{u0}<0$ and $r_{v0}>0$ in order to make the radial function for an ingoing observer decreasing and that for an outgoing observer increasing.

When we compare the double-null coordinate with the static solution $ds^{2} = - N^{2}(r) dt^{2} + dr^{2} /N^{2}(r) + r^{2} d\Omega^{2}$, the following relations are useful:
\begin{eqnarray}
dr = r_{,u}du + r_{,v}dv,\;\;\;\;\; dt = \frac{\alpha^{2}}{4} \left( -\frac{dv}{r_{,u}} + \frac{du}{r_{,v}} \right),\;\;\;\;\;
N^{2} = - \frac{4 r_{,u} r_{,v}}{\alpha^{2}}.
\end{eqnarray}
If we present a static solution $N^{2} = 1 - 2 m/r + Q^{2}/r^{2} - \Lambda r^{2}/3$ with mass $m$, charge $Q$, and the cosmological constant or vacuum energy $\Lambda$, we obtain \cite{Hwang:2011mn}
\begin{eqnarray}
m(u,v) = \frac{r}{2} \left( 1 + \frac{4 r_{,u} r_{,v}}{\alpha^{2}} + \frac{Q^{2}}{r^{2}} - \frac{\Lambda}{3} r^{2} \right).
\end{eqnarray}
By using this, one can give a consistent boundary condition for generic situations.

\begin{description}
\item[Ingoing null surface:]
In order to derive a gravitational collapse, it would be convenient to use a shell-shaped scalar field, where its interior is identified with Minkowski. Then it is convenient to choose $r_{u0}=-1/2$ and $r_{v0}=1/2$. We need to choose the mass function on $u_{\mathrm{i}}=v_{\mathrm{i}}=0$ to vanish, and hence, $\alpha(0,0) = 1$. At the same time, $s(u,0) = w(u,0) = 0$. By plugging this into Eq. (\ref{eq:E1}), we obtain $h(u,0) = 0$. Finally, we need more information to determine $d, g$, and $z$ on the $v=0$ surface. We obtain $d$ from Eq. (\ref{eq:E4}), $g$ from Eq. (\ref{eq:E3}), and $z$ from Eq. (\ref{eq:s}).

\item[Outgoing null surface:] We can choose an arbitrary function for $s(0,v)$ to induce a collapsing pulse, for example,
\begin{eqnarray}
s(0,v)= A \sin^{2} \left( \pi \frac{v-v_{\mathrm{i}}}{v_{\mathrm{f}}-v_{\mathrm{i}}} \right)
\end{eqnarray}
for $v_{\mathrm{i}}\leq v \leq v_{\mathrm{f}}$ and otherwise $s(0,v)=0$. Note that the initial condition must be continuous up to first derivatives of $v$. From this, we obtain $z(0,v)=s(0,v)_{,v}$. From Eq. (\ref{eq:E2}), since $g_{,v}(0,v)=0$, we obtain $d(0,v)$. By integrating $d$ along $v$, we have $\alpha(0,v)$. Finally, we need more information for $h, f$, and $w$ on the $u=0$ surface. We obtain $h$ from Eq. (\ref{eq:E4}), $f$ from Eq. (\ref{eq:E3}), and $w$ from Eq. (\ref{eq:s}).
\end{description}
This procedure is summarized in Table \ref{table:conditions}.

\begin{table}[ph]
\tbl{\label{table:conditions}Summary of the assignments of initial conditions for $v=v_{\mathrm{i}}$ and $u=u_{\mathrm{i}}$.}
{\begin{tabular}{@{}cccc@{}}
\hline
& \;\;fixed by hand\;\; & \;\;fixed by constraints\;\; & \;\;fixed by evolution equations\;\;\\
\hline \hline
$v=v_{\mathrm{i}}$ & $s = 0, f, r$ & $w=0$, $h=0$, $\alpha=1$ & $d, g, z$\\
\hline
$u=u_{\mathrm{i}}$ & $s, g, r$ & $z=s_{,v}$, $d=rz^2$, $\alpha \leftarrow d$ & $h, f, w$\\
\hline
\end{tabular}}
\end{table}

\subsection{Solving algorithm}

There are several possibilities for prescribing a numerical algorithm to solve the field equations introduced in Sec. \ref{sec:fieldeqns}. One possibility is the one proposed in \cite{cqg13(1996)497}. The computations are conducted on a two-dimensional numerical grid constructed within the $(uv)$-plane. A value of a particular function at a certain point $\left(u,v\right)$ stems from values of the appropriate functions at points $\left(u,v-\Delta_v\right)$ and $\left(u-\Delta_u,v\right)$, with $\Delta_v$ and $\Delta_u$ being integration steps in the ingoing and outgoing directions, respectively. The differential field equations of the introduced variables along $u$ and $v$ can be symbolically written as
\begin{equation}
\textsf{f}_{,u} \ \; = \ \; \textsf{F}\left(\textsf{f},\textsf{g}\right), \qquad
\textsf{g}_{,v} \ \; = \ \; \textsf{G}\left(\textsf{f},\textsf{g}\right).
\end{equation}
The values of the unknown quantities are given by
\begin{eqnarray}
\textsf{f}\big\arrowvert_{\left(u,v\right)} &=& \frac{1}{2}\bigg(\textsf{ff}\big\arrowvert_{\left(u,v\right)}+
\textsf{f}\big\arrowvert_{\left(u-\Delta_u,v\right)} +
\Delta_u\textsf{F}\left(\textsf{ff},\textsf{gg}\right)\big\arrowvert_{\left(u,v\right)}\bigg), \\
\textsf{g}\big\arrowvert_{\left(u,v\right)} &=& \frac{1}{2}\bigg(\textsf{gg}\big\arrowvert_{\left(u,v\right)}+
\textsf{g}\big\arrowvert_{\left(u,v-\Delta_v\right)} +
\Delta_v\textsf{G}\left(\textsf{ff},\textsf{gg}\right)\big\arrowvert_{\left(u,v\right)}\bigg),
\end{eqnarray}
where the auxiliary quantities are
\begin{eqnarray}
\textsf{ff}\big\arrowvert_{\left(u,v\right)} &=& \textsf{f}\big\arrowvert_{\left(u-\Delta_u,v\right)} + 
\Delta_u\textsf{F}\left(\textsf{f},\textsf{g}\right)\big\arrowvert_{\left(u-\Delta_u,v\right)}, \\
\textsf{gg}\big\arrowvert_{\left(u,v\right)} &=& \textsf{g}\big\arrowvert_{\left(u,v-\Delta_v\right)} +
\frac{\Delta_v}{2}\bigg(\textsf{G}\left(\textsf{f},\textsf{g}\right)\big\arrowvert_{\left(u,v\right)} +
\textsf{G}\left(\textsf{ff},\textsf{gg}\right)\big\arrowvert_{\left(u,v\right)}\bigg).
\end{eqnarray}

\begin{figure}
\begin{center}
\includegraphics[scale=0.3]{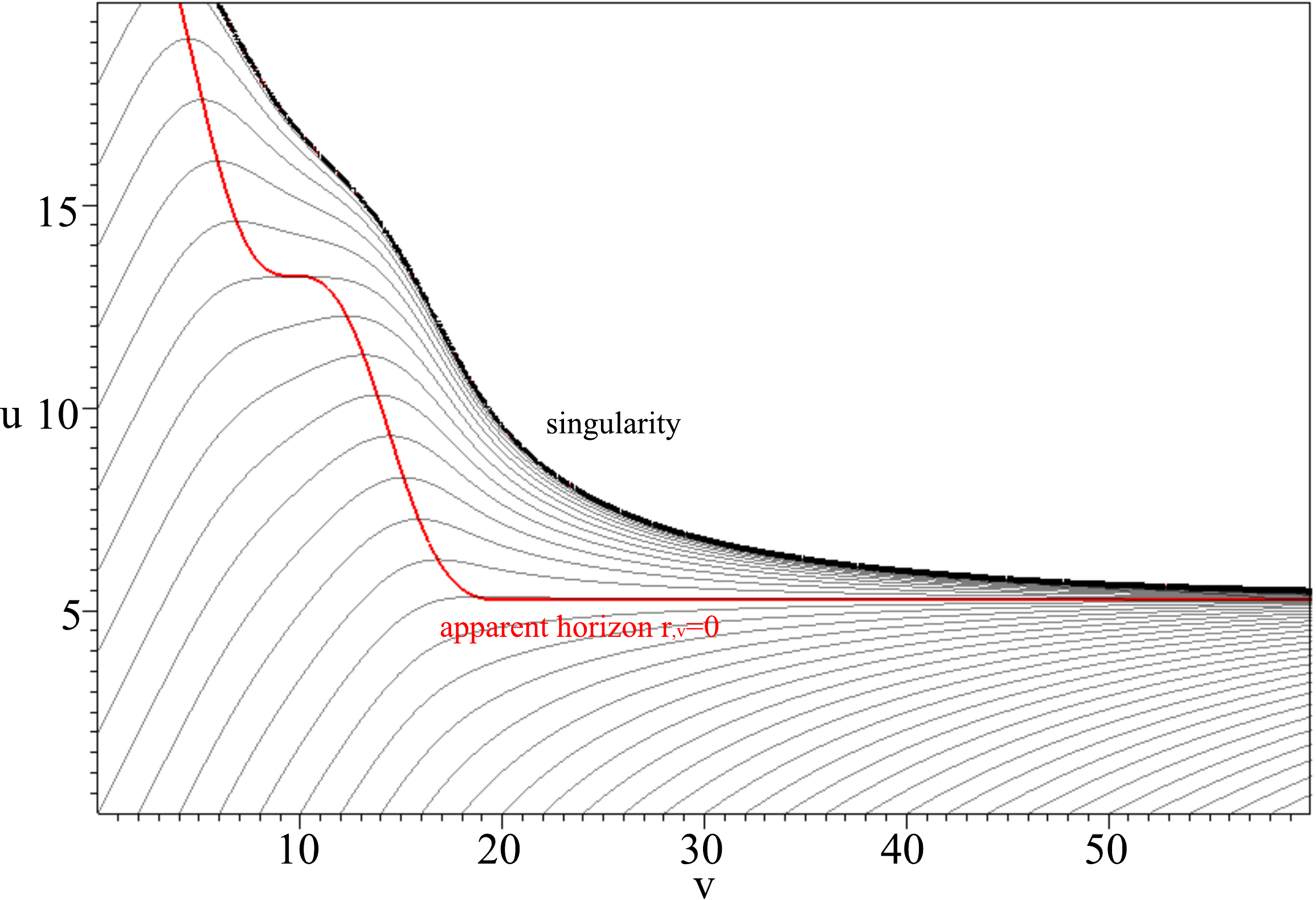}
\caption{\label{fig:neutral}An example of a numerical simulation with $A = 0.5$. Thin curves are contours of constant $r$, a thick black curve is the singularity $r = 0$, and a thin red curve is the apparent horizon $r_{,v} = 0$.}
\includegraphics[scale=0.7]{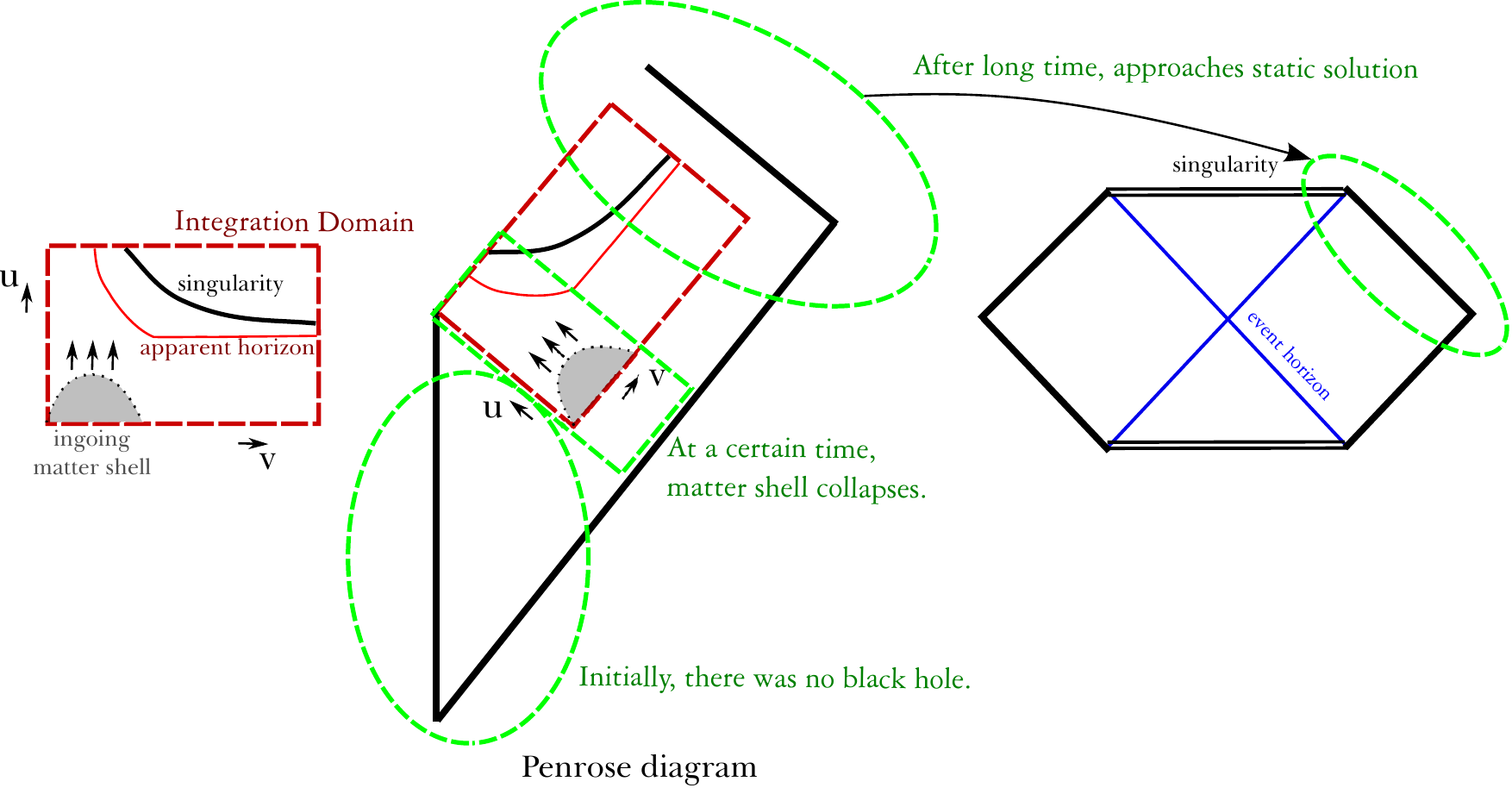}
\caption{\label{fig:domain}Left: Integration domain of simulations. We obtain a two-dimensional data. Middle: Since $u$ and $v$ directions are null, by tilting $45$-degree, we obtain a Penrose diagram. Right: By giving proper boundary conditions, one may assume that initially there was no black hole. After gravitational collapse finished, one can match the solution to the Penrose diagram of a Schwarzschild black hole.}
\end{center}
\end{figure}

The double-null coordinates ensure regular behavior of all the variables within the domain of integration except the vicinity of $r=0$, where the singularity resides. During a numerical analysis difficulties also arise in areas, where the function $f$ 
diverges. A relatively dense numerical grid is necessary in order to conduct the computations in these regions.
The efficiency of the calculations suggests using an adaptive grid which allows to perform integration with an appropriately smaller step in particular regions, where it is needed. A refinement algorithm making the grid denser only along the ingoing direction is sufficient. A local error indicator related to the variables and changing its value significantly in the adequate region is the function ${\Delta r}/r$ with the difference in $r$, $\Delta r$, calculated along the $u$-coordinate \cite{prd68(2003)044013,Thornburg:2009mw}. In general, one can also apply the adaptive mesh for both of $u$ and $v$ directions \cite{Burko:1998az}. In addition, it is also equivalently useful to adaptively choose the gauge degrees of freedom for several circumstances \cite{Eilon:2015axa}.

\subsection{Sample results}

For a numerical simulation, we can choose $v_{\mathrm{f}}=20$ and leave $A$ as a free parameter (Fig. \ref{fig:neutral}) \cite{Hong:2008mw}. This shows a dynamical formation of a black hole, including a formation of an apparent horizon $r_{,v} = 0$ as well as a singularity ($r = 0$). Based on this numerical result, finally we interpret the results as in Fig. \ref{fig:domain} \cite{Hwang:2013zaa}. We obtain numerical results for a given integration domain $(u=0,u=u_{\mathrm{max}}) \times (v=0, v=v_{\mathrm{max}})$ (left). Since $u$ and $v$ directions are null, we can easily read off the structire of a Penrose diagram. By giving proper boundary conditions, one may assume that initially there was no black hole. After gravitational collapse finished, one can match the solution to the Penrose diagram of a Schwarzschild black hole (right).

\section{\label{sec:ext}Extensions}

In this section, we review possible extensions of this double-null formalism.

\subsection{Modifying matter}

The most trivial way to extend the theory is to change the matter sector.

\subsubsection{$U(1)$ gauge field} One can add a complex-scalar field with a $U(1)$ gauge symmetry in order to introduce electric charges by considering the following Lagrangian density\cite{Hong:2008mw,prd68(2003)044013,Hod:1998gy,Hansen:2005am,Burko:1997tb,Burko:1997zy}:
\begin{eqnarray}\label{eqn:gf}
\mathcal{L}_{\mathrm{EM}} = - \frac{1}{2}\left(\phi_{;\mu}+ieA_{\mu}\phi \right)g^{\mu\nu}\left(\overline{\phi}_{;\nu}-ieA_{\nu}\overline{\phi}\right)-\frac{1}{16\pi}F_{\mu\nu}F^{\mu\nu},
\end{eqnarray}
where $\phi$ is a complex scalar field with a complex conjugate denoted with a bar, $A_{\mu}$ is a gauge field, $e$ is their coupling, and $F_{\mu\nu} = A_{\nu;\mu} - A_{\mu;\nu}$. In the double-null formalism, one can fix a gauge such that $A_{\mu} = [A_{u},0,0,0]$. Then, one can prove that the corresponding charge is $q \equiv 2 r^{2} A_{u,v}/\alpha^{2}$. The relevant components of the energy-momentum tensor are
\begin{eqnarray}
&&T_{uu} = \frac{1}{4\pi} \left[ w\bar{w} + ieA_u \left( \bar{w} s - w\bar{s} \right) + e^2A_u^2s\bar{s} \right], \;\;\;\;\; T_{vv} = \frac{1}{4\pi} z\bar{z}, \\
&&T_{uv} = \frac{A_{u,v}^2}{4\pi\alpha^2}, \;\;\;\;\; 
T_{\theta\theta} = \frac{r^2}{4\pi\alpha^2}
\left[ w\bar{z} + z\bar{w} + ieA_u \left( \bar{z} s - z\bar{s} \right) + \frac{2A_{u,v}^2}{\alpha^2} \right],
\end{eqnarray}
and field equations are
\begin{eqnarray}
&& A_{u,v} = \frac{\alpha^2 q}{2r^2}, \;\;\;\;\; q_{,v} = -\frac{ier^2}{2} \left( \bar{s} z-s\bar{z} \right), \\
&& z_{,u} =w_{,v} = -\frac{fz}{r} - \frac{gw}{r} - ieA_u z - \frac{ieA_u gs}{r} - \frac{ie}{4r^2}\alpha^2 qs.
\end{eqnarray}

\begin{figure}
\begin{center}
\includegraphics[scale=0.27]{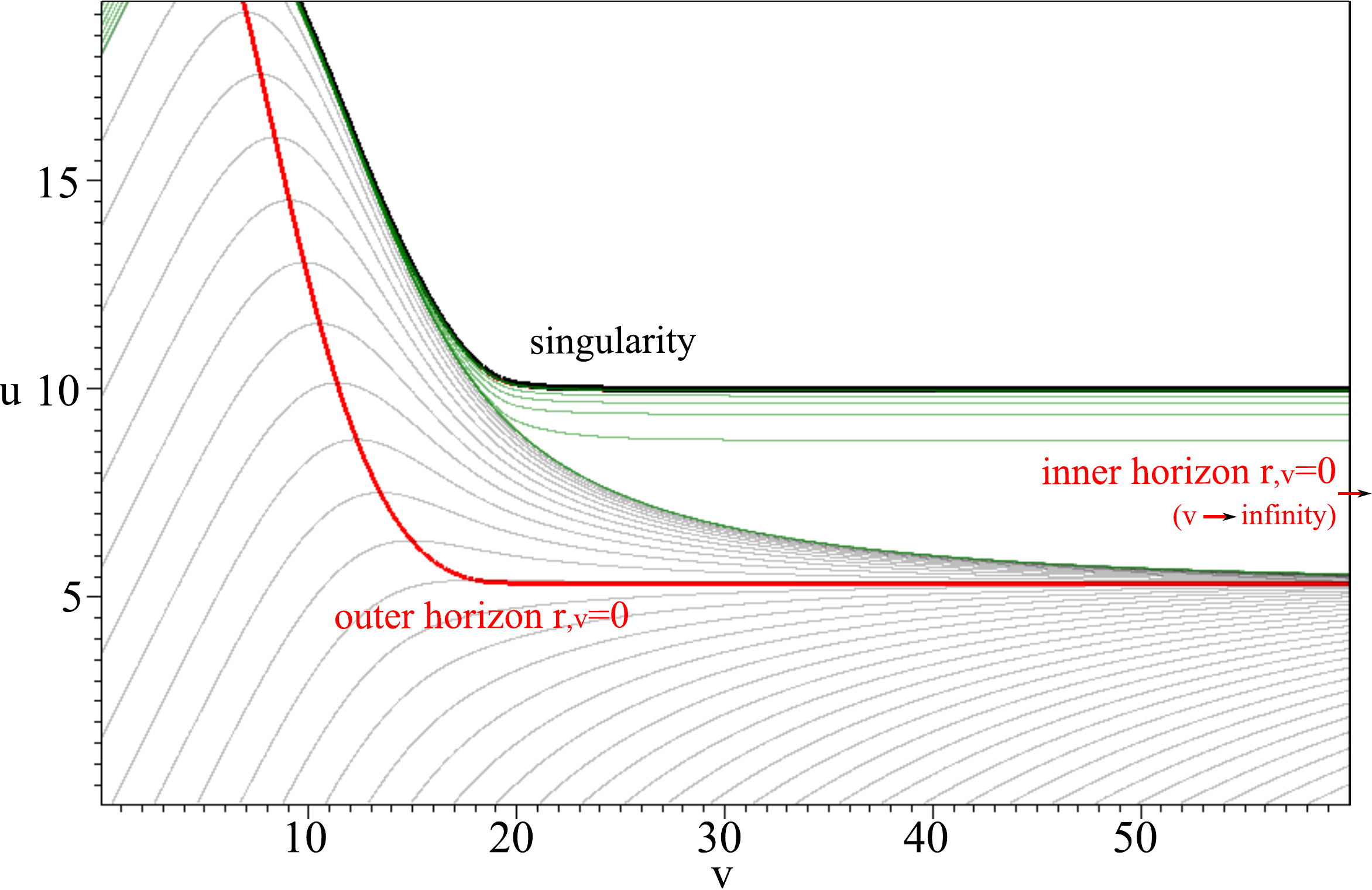}
\includegraphics[scale=0.26]{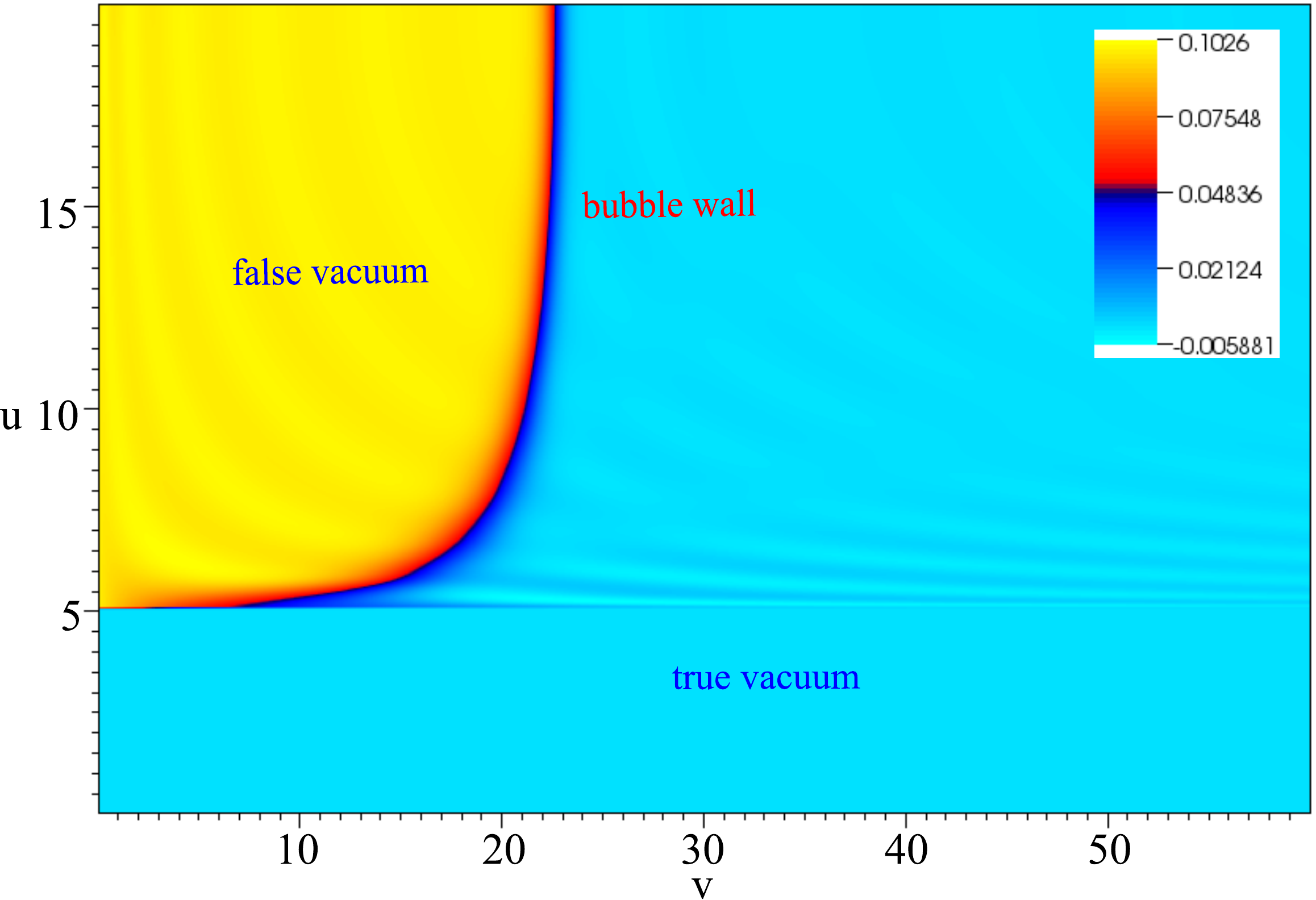}
\caption{\label{fig:bubble}Left: an example of a charged black hole with $e=0.3$ and $A=0.25$, where the spacing is $1$ for black contours and $0.1$ for green contours. Right: an example of a collapsing false vacuum bubble, where the color corresponds to $\phi(u,v)$ and the yellow colored part is a false vacuum and the blue colored part is a true vacuum.}
\end{center}
\end{figure}

One another essential point is to give a proper initial condition. The electric charge will increase as the real part and imaginary part of $\phi$ has a non-trivial phase difference. One simple example is
\begin{eqnarray}
s(0,v)= A \sin^{2} \left( \pi \frac{v-v_{\mathrm{i}}}{v_{\mathrm{f}}-v_{\mathrm{i}}} \right) \exp\left(2\pi i \frac{v - v_{\mathrm{i}}}{v_{\mathrm{f}} - v_{\mathrm{i}}} \right)
\end{eqnarray}
for $v_{\mathrm{i}} \leq v \leq v_{\mathrm{f}}$, while one may choose more complicated forms. See an example on the left panel of Fig. \ref{fig:bubble}\cite{Hong:2008mw}.

\subsubsection{Vacuum energy and bubbles} One can include a potential term to a scalar field, where it can mimic an inflaton field\cite{Hansen:2009kn,Hwang:2010gc}:
\begin{eqnarray}
\mathcal{L}_{\mathrm{V}} = - \frac{1}{2} \phi_{;\mu} \phi_{;\nu} g^{\mu\nu} - V(\phi),
\end{eqnarray}
where $V(\phi)$ is an arbitrary potential. The $uv$ and $\theta\theta$ components of the energy-momentum tensor and the Klein-Gorden equation of a real scalar field will be modified:
\begin{eqnarray}
T_{uv} = \frac{\alpha^{2}}{2} V, \;\;\; T_{\theta\theta} = \frac{r^{2}}{2\pi\alpha^{2}} wz - r^{2} V, \;\;\; z_{,u} = w_{,v} = - \frac{fz}{r} - \frac{gw}{r} - \pi \alpha^{2} V'(s),
\end{eqnarray}
where $V(s) = V(\phi)|_{\phi=s/\sqrt{4\pi}}$.

In order to describe vacuum bubbles, one can choose the following initial condition for an outgoing bubble:
\begin{eqnarray}
\phi(u,v_{\mathrm{i}}) = \left\{ \begin{array}{ll}
0 & u < u_{\mathrm{shell}},\\
\phi_{0} G(u) & u_{\mathrm{shell}} \leq u < u_{\mathrm{shell}}+\Delta u,\\
\phi_{0} & u_{\mathrm{shell}}+\Delta u \leq u,
\end{array} \right.
\end{eqnarray}
where $G(u)$ is a pasting function such as $G(u) = \sin^{2} \left[\pi(u-u_{\mathrm{shell}})/2\Delta u\right]$. Here, $V(\phi_{0})$ and $V_{0}$ correspond to a true or false vacuum, $u_{\mathrm{shell}}$ denotes the location of the shell, and $\Delta u$ is the thickness of the shell. The same thing can be applied for an ingoing shell by changing $u$ and $v$. It is possible that if the kinetic term of the scalar field is opposite, then one can even build an inflating bubble. See an example on the right panel of Fig. \ref{fig:bubble}\cite{Hansen:2009kn}.

\subsubsection{Phantom field}

One possible modification of the matter sector which leads to interesting dynamical spacetime structures is to consider a phantom coupling of the scalar field. Such a phantom field has a sign in front of the kinetic term in the Lagrangian opposite to the standard one. In the case of a phantom counterpart of the scalar field discussed in Sec. \ref{sec:the}, the collapse does not lead to any singular structures. When a phantom version of the $U(1)$ gauge field described by (\ref{eqn:gf}) is subjected to gravitational evolution, the Schwarzschild spacetime emerges, instead of a Reissner-Nordstr\"om one.

\begin{figure}
\begin{center}
\includegraphics[scale=0.17]{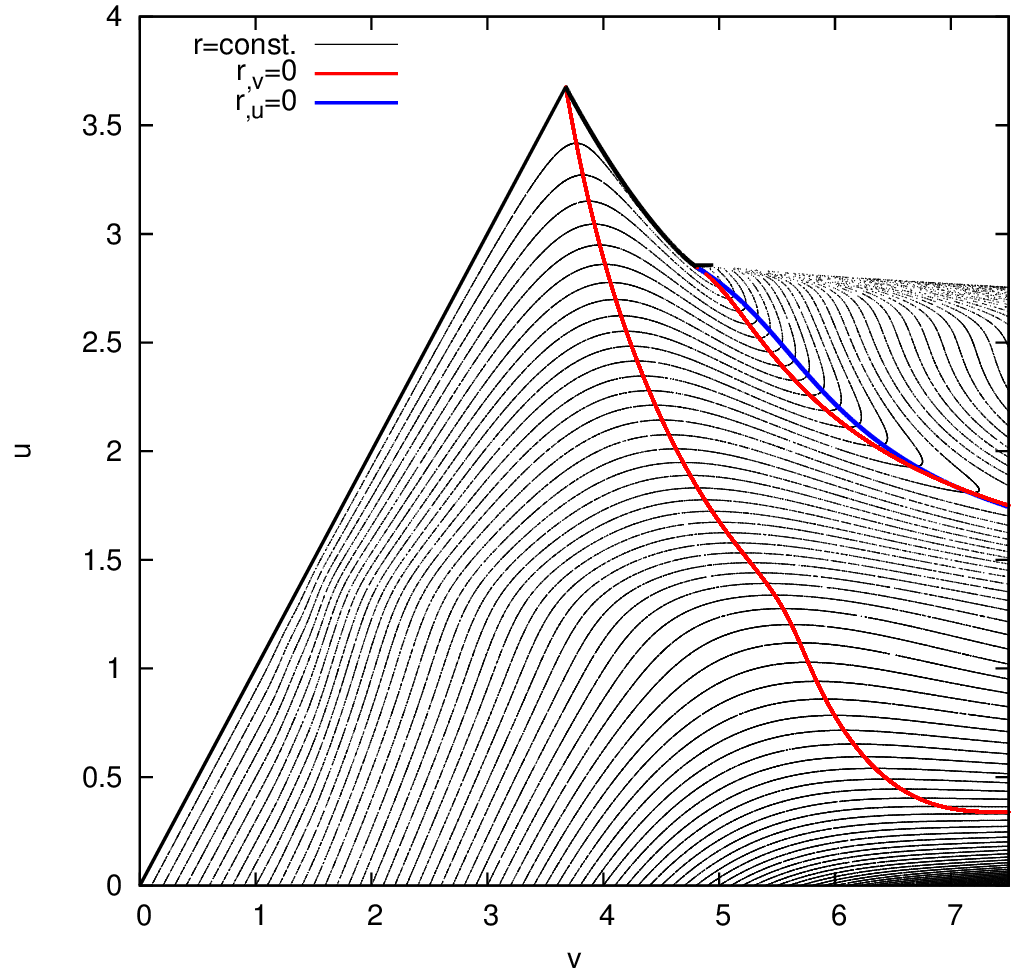} \hspace{0.5cm} \includegraphics[scale=0.3]{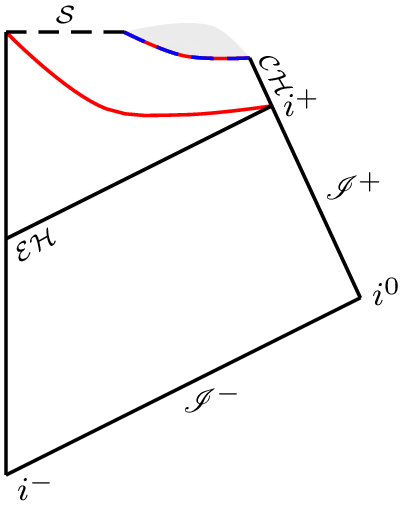}
\caption{\label{fig:worm}An example of a dynamical wormhole. Left: spacetime diagram with two types of apparent horizons ($r_{,v}=0$ and $r_{,u}=0$ as red and blue curves, respectively). The singularity is depicted as a thick black curve. Right: The corresponding Penrose diagram with the event (EH) and Cauchy (CH) horizons depicted, along with the central singularity (S).}
\end{center}
\end{figure}

\begin{figure}
\begin{center}
\includegraphics[scale=0.17]{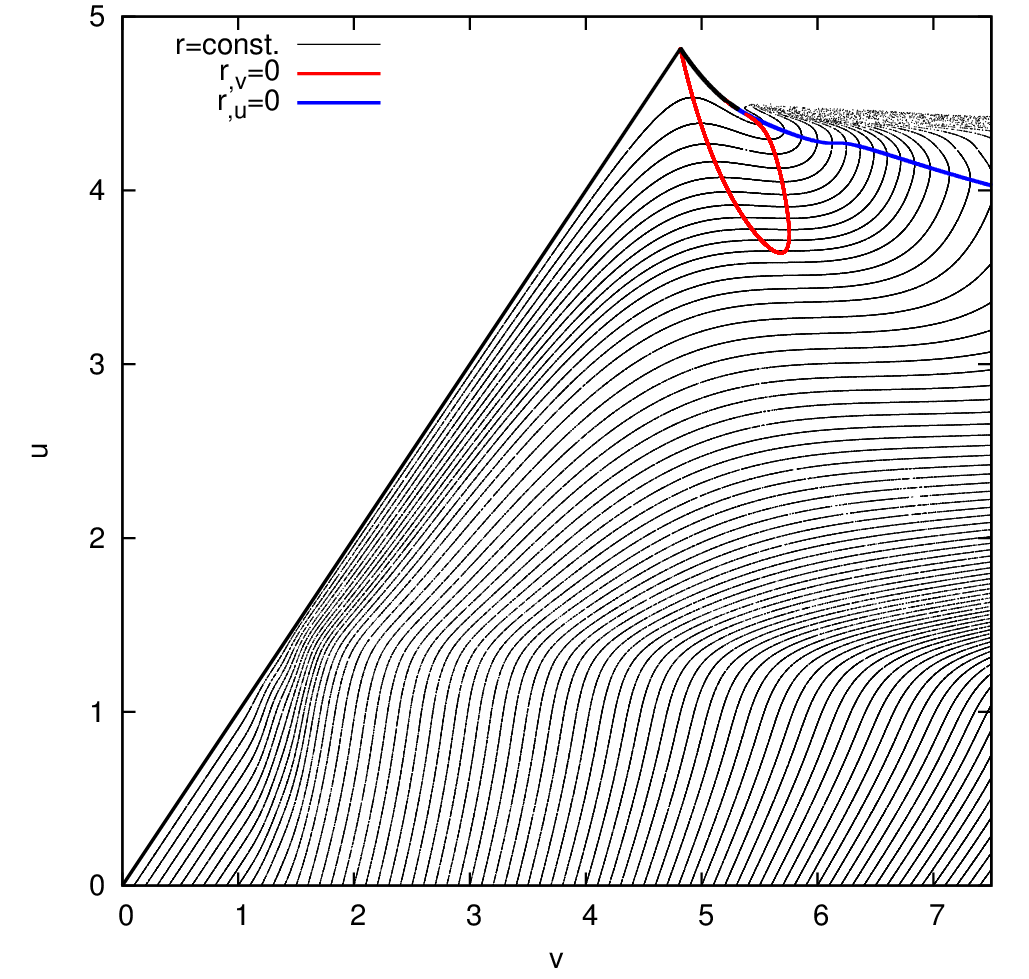} \hspace{0.5cm} \includegraphics[scale=0.3]{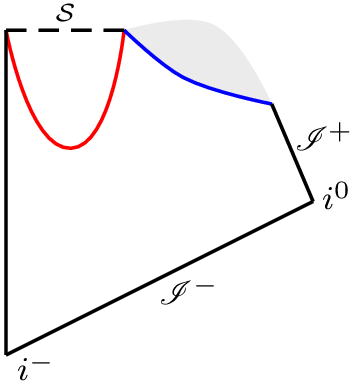}
\caption{\label{fig:ns}An example of a dynamical naked singularity. Left: spacetime diagram with two types of apparent horizons ($r_{,v}=0$ and $r_{,u}=0$ as red and blue curves, respectively). The singularity is depicted as a thick black curve. Right: The corresponding Penrose diagram with the singularity denoted as $S$.}
\end{center}
\end{figure}

More complex and exotic spacetime objects are formed during evolutions involving more than one matter types, one of which is phantom. Figs. \ref{fig:worm} and \ref{fig:ns} show specetimes containing a dynamical wormhole and a naked singularity, respectively. They stem from the collapse of a phantom scalar field $\psi$ accompanied by another scalar field charged under a $U(1)$ gauge field, with the overall setup within a low-energy limit of the string theory. In the string frame, the adequate Lagrangian is \cite{prd83(2011)084007,grg44(2012)3175,prd86(2012)044043}
\begin{eqnarray}
\mathcal{L}_{\mathrm{PH}} = e^{-2\psi} \left[ R + 2\left( \nabla\psi \right)^2 + e^{-2\psi} \mathcal{L}_{EM} \right].
\end{eqnarray}

\begin{figure}
\begin{center}
\includegraphics[scale=0.3]{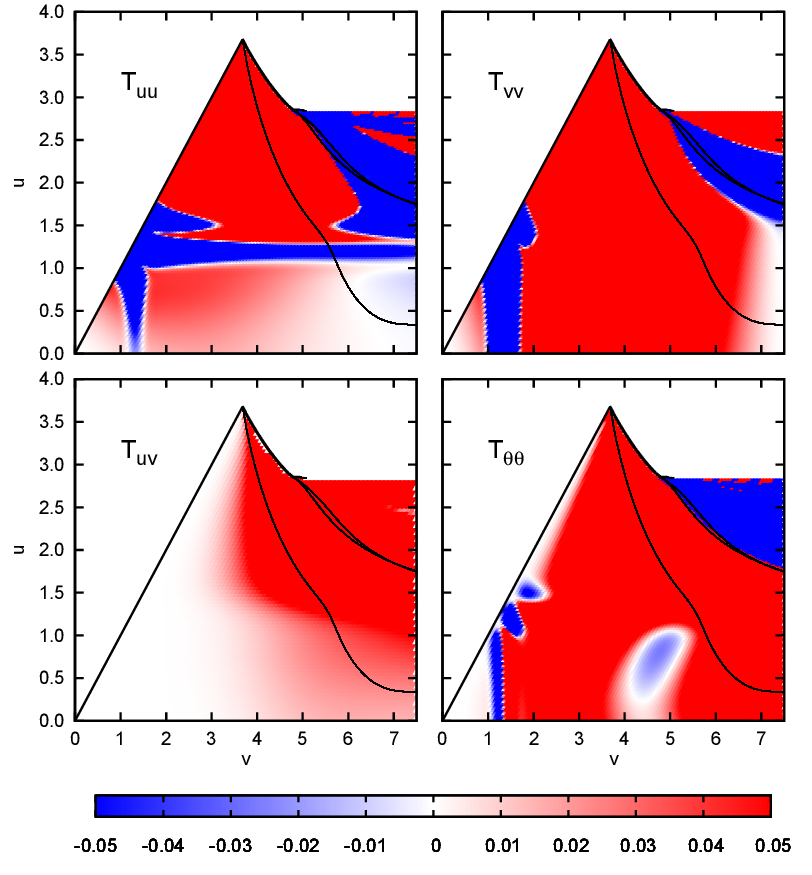}
\caption{\label{fig:setc}The stress-energy tensor components within the spacetime containing a dynamical wormhole presented on the left panel of Fig. \ref{fig:worm}.}
\end{center}
\end{figure}

Phantom fields, due to their gravitationally repulsive character, seem to be capable of fostering spacetime elements which are of a milder character than a strong singularity. Among these are wormhole throats, beyond which the spacetime extends, or naked singularities, surrounded only by an apparent horizon, not an event horizon, and not spanning to infinity within spacetime. This feature may be linked to the fact that phantom fields violate the null energy condition (NEC) $T_{\mu\nu}n^\mu n^\nu \geqslant 0$, with $n^\mu$ being a null vector, which in spherical symmetry and double-null coordinates is equivalent to $T_{uu}\geqslant 0$ and $T_{vv}\geqslant 0$. The NEC violation appears in the vicinity of wormhole throats, as can be inferred from Fig. \ref{fig:setc}.

\begin{figure}
\begin{center}
\includegraphics[scale=0.27]{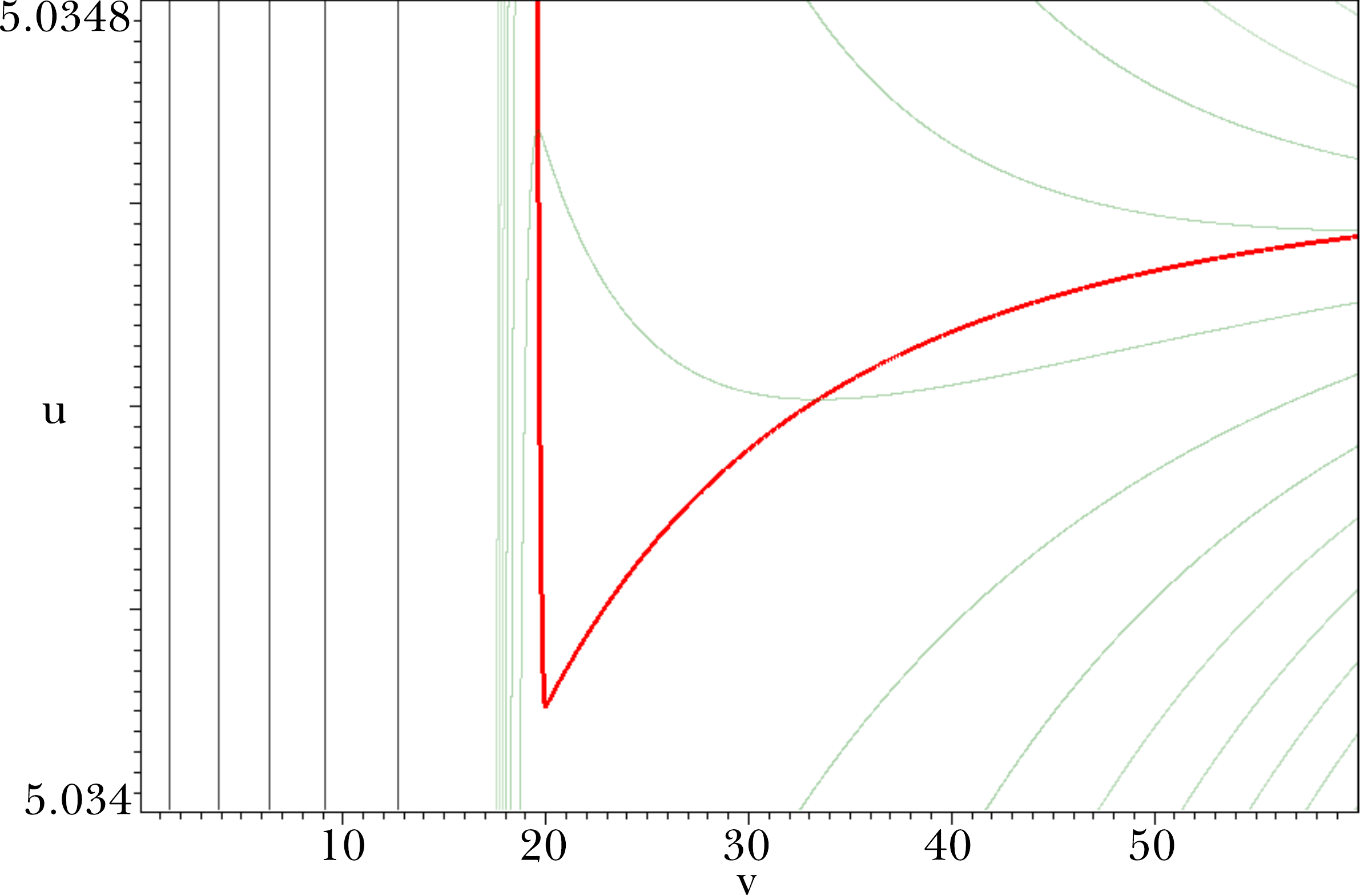}
\caption{\label{fig:Schwarzschild}An example of an evaporating black hole, where the spacing is $1$ for black contours and $0.002$ for green contours. The apparent horizon ($r_{,v}=0$, red colored curve) is bending to a time-like direction.}
\end{center}
\end{figure}

\subsubsection{Quantum effects: Hawking radiation} One can even consider semi-classical effects, at least by an approximate form. By using the $S$-wave approximation, we use the exact two-dimensional calculation for $T_{\mu\nu}$\cite{Davies:1976ei} (up to one loop order) divided by $4\pi r^{2}$ and we obtain
\begin{equation}
\langle \hat{T}_{uu} \rangle = \frac{P \left(h_{,u}-h^{2}\right)}{4\pi r^{2}},\;\;\;\; \langle \hat{T}_{uv} \rangle = \langle \hat{T}_{vu} \rangle = -\frac{P}{4\pi r^{2}}d_{,u},\;\;\;\; \langle \hat{T}_{vv} \rangle = \frac{P\left(d_{,v}-d^{2}\right)}{4\pi r^{2}},
\end{equation}
with $P \equiv Nl_{\mathrm{Pl}}^2 / 12\pi$, where $N$ is the number of massless scalar fields and $l_{\mathrm{Pl}}$ is the Planck length. Then we need to solve the semi-classical Einstein equation
\begin{eqnarray}
G_{\mu\nu}=8\pi \left( T_{\mu\nu}+\langle \hat{T}_{\mu\nu} \rangle \right).
\end{eqnarray}
One interesting observation is that due to this $P$-dependent term, the equation for $\alpha_{,uv}$ is (for a free scalar field model)
\begin{eqnarray}
d_{,u} = h_{,v} &=& \frac{1}{1-P/r^{2}} \left[ \frac{fg}{r^{2}} + \frac{\alpha^2}{4r^{2}} - wz \right].
\end{eqnarray}
Hence, there appears a semi-classical singularity at $r = \sqrt{P} \simeq \sqrt{N} l_{\mathrm{Pl}}$ which is the same as that of Dvali's semi-classical cutoff \cite{Dvali:2007hz}. There are several papers that show this approximation gives a good result for evaporating black holes\cite{Piran:1993tq,Parentani:1994ij,Ayal:1997ab} (Fig. \ref{fig:Schwarzschild})\cite{Hong:2008mw}. Note that if we increase $N$ by fixing $P$, then this means that the simulation is the same since we fixed a simulation parameter, but the interpretation in terms of semi-classical physics is changed and the unit length becomes smaller and smaller; all curvature quantities become smaller and smaller in terms of the Planck scales\cite{Yeom:2009zp}.

\subsection{Modifying gravity}

One can also consider modified gravity. The prototype model is\cite{Hansen:2014rua,Hansen:2015dxa,jhep02(2016)049,jhep05(2016)155}
\begin{eqnarray}
S_{\mathrm{BD}} = \int d^4 x \sqrt{-g} \left[ \frac{1}{16\pi} 
\left( \Phi R - \frac{\omega}{\Phi} \Phi_{;\mu} \Phi_{;\nu} g^{\mu\nu} \right) + \Phi^\beta \mathcal{L}_{\mathrm{M}} \right],
\end{eqnarray}
where $\Phi$ denotes the Brans-Dicke field, $\omega$ is the Brans-Dicke coupling constant and $\beta$ is a model-dependent constant which will be explained later. The Einstein equations can be written as
\begin{eqnarray}
G_{\mu\nu} = 8\pi \left( T_{\mu\nu}^{\mathrm{BD}} + \Phi^{\beta-1} T_{\mu\nu}^{\mathrm{M}} \right),
\end{eqnarray}
where
\begin{eqnarray}
T_{\mu\nu}^{\mathrm{BD}} = \frac{1}{8\pi\Phi} \left( \Phi_{;\mu\nu} - g_{\mu\nu}\Phi_{;\rho\sigma} g^{\rho\sigma} \right)
+ \frac{\omega}{8\pi\Phi^2} \left( \Phi_{;\mu} \Phi_{;\nu} - \frac{1}{2} g_{\mu\nu} \Phi_{;\rho} \Phi_{;\sigma} g^{\rho\sigma} \right).
\end{eqnarray}
The equation of motion of the Brans-Dicke field is
\begin{eqnarray}
\Phi_{;\mu\nu} g^{\mu\nu} - \frac{8\pi\Phi^\beta}{3+2\omega} \left( T^{\mathrm{M}} - 2\beta \mathcal{L}_{\mathrm{M}} \right) = 0,
\end{eqnarray}
where $T^{\mathrm{M}} = {T^{\mathrm{M}}}^{\mu}_{\;\mu}$.

\begin{figure}
\begin{center}
\includegraphics[scale=0.23]{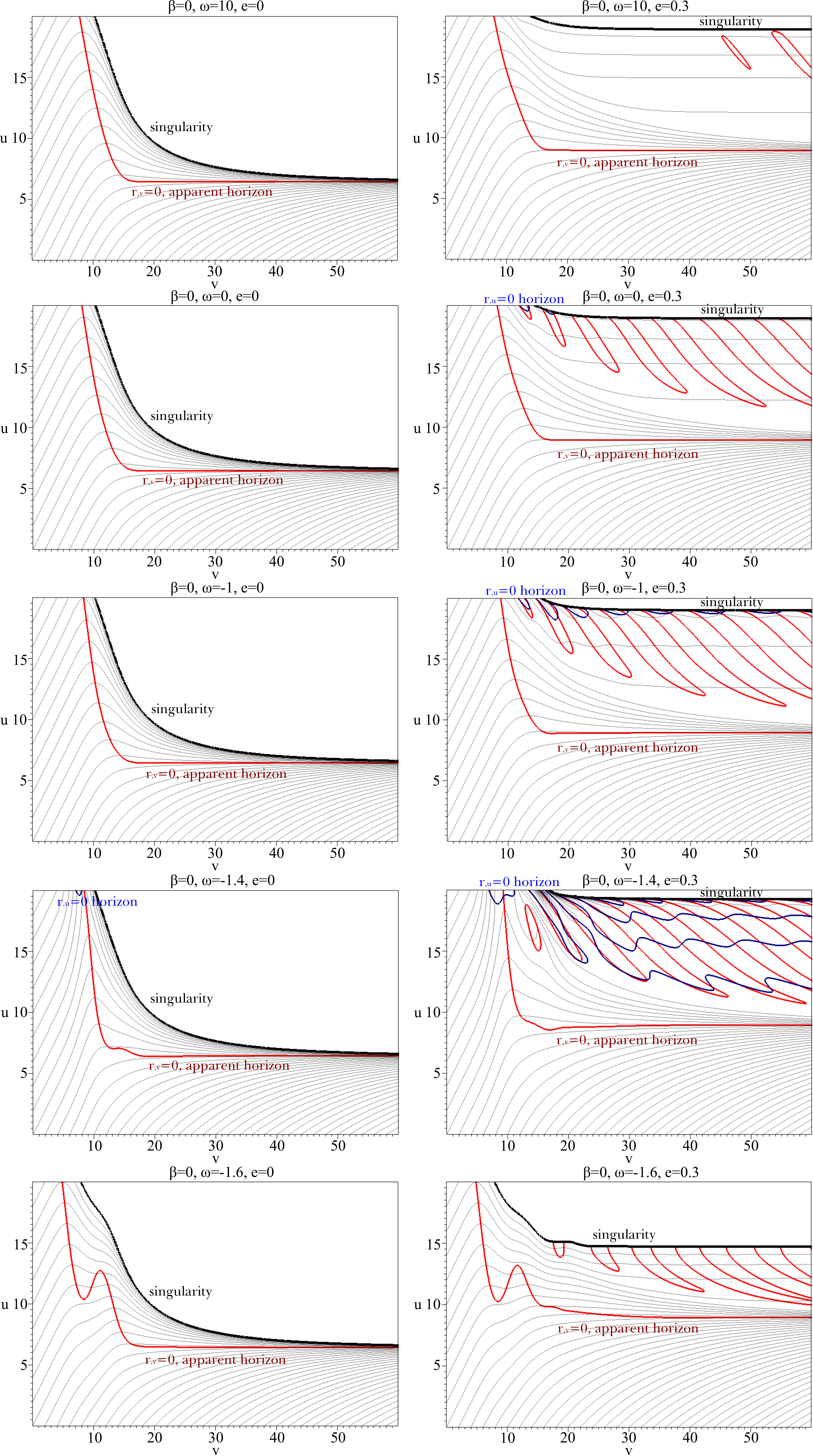}
\caption{\label{fig:beta=0}An example of simulations for $\beta = 0$ with various $\omega$ and $e$.}
\end{center}
\end{figure}

The energy-momentum tensor components are
\begin{eqnarray}
T_{uu}^{\mathrm{BD}} &=& \frac{1}{8\pi\Phi} \left( W_{,u} - 2hW \right) + \frac{\omega}{8\pi\Phi^2}W^2, \\
T_{vv}^{\mathrm{BD}} &=& \frac{1}{8\pi\Phi} \left( Z_{,v} - 2dZ \right) + \frac{\omega}{8\pi\Phi^2}Z^2, \\
T_{uv}^{\mathrm{BD}} &=& -\frac{Z_{,u}}{8\pi\Phi} - \frac{gW + fZ}{4\pi r\Phi}, \\
T_{\theta\theta}^{\mathrm{BD}} &=& \frac{r^2}{2\pi\alpha^2 \Phi}Z_{,u}
+ \frac{r}{4\pi\alpha^2 \Phi} \left( gW + fZ \right) + \frac{\omega r^2}{4\pi\Phi^2\alpha^2}WZ,
\end{eqnarray}
where $\Phi_{,u} \equiv W$ and $\Phi_{,v} \equiv Z$. In addition, it is convenient to introduce the notation $\tilde{A} \equiv \Phi^{\beta} A$ with an arbitrary function $A$. Then, the equations for $\alpha_{,uv}$, $r_{,uv}$, and $\Phi_{,uv}$ may be gathered as follows:
\begin{eqnarray}
\begin{pmatrix}
d_{,u} = h_{,v} \\
r_{,uv} \\
\Phi_{,uv}
\end{pmatrix}
= \frac{1}{r^2}
\begin{pmatrix}
r^2\ & -r\ & -\frac{r}{2\Phi} \\
0\ & r^2\ & -\frac{r^2}{2\Phi} \\
0\ & 0\ & r
\end{pmatrix}
\begin{pmatrix}
\mathcal{A} \\
\mathcal{B} \\
\mathcal{C}
\end{pmatrix},
\end{eqnarray}
where
\begin{eqnarray}
\mathcal{A} &\equiv& -\frac{2\pi\alpha^2}{r^2\Phi} \widetilde{T}_{\theta\theta}^{\mathrm{M}} - \frac{1}{2r\Phi} \left( gW+fZ \right)
- \frac{\omega}{2\Phi^2}WZ, \\
\mathcal{B} &\equiv& -\frac{\alpha^2}{4r} - \frac{fg}{r} + \frac{4\pi r}{\Phi} \widetilde{T}_{uv}^{\mathrm{M}}
- \frac{1}{\Phi} \left( gW+fZ \right), \\
\mathcal{C} &\equiv& -fZ -gW - \frac{2\pi r\alpha^2}{3+2\omega} \left( \widetilde{T}^{\mathrm{M}} - 2\beta\widetilde{\mathcal{L}}_{\mathrm{M}} \right).
\end{eqnarray}

Examples of the simulation results are shown in Fig. \ref{fig:beta=0}\cite{jhep02(2016)049,jhep05(2016)155}. We comment on theoretical motivations of the model parameters in the following sections.

\subsubsection{Dilaton-matter couplings}

The prototype model introduced above can have several applications\cite{Gasperini:2007zz}. First, this can be viewed as presenting a kind of dilaton-matter couplings. The expansions of the effective actions of the bosonic sector of the type IIA, type I and heterotic string theories are
\begin{eqnarray}
S_{\textrm{IIA}} &=& \frac{1}{2\lambda_s^{8}} \int d^{10}x \sqrt{-g} \Bigg\{ e^{-\phi_d} \left[ R + \big( \nabla\phi_d \big)^2 - \frac{H_3^{2}}{12} \right] - \left( \frac{F_2^{2}}{4} + \frac{\tilde{F}_4^{2}}{48} \right) \Bigg\} 
+ \cdots, \nonumber \\
S_{\textrm{I}} &=& \frac{1}{2\lambda_s^{8}} \int d^{10}x \sqrt{-g} \Bigg\{ e^{-\phi_d} \left[ R + \big( \nabla\phi_d \big)^2 \right] - \frac{\tilde{H}_3^{2}}{12} - e^{-\frac{\phi_d}{2}} \frac{\textrm{Tr}\big(F_2^{2}\big)}{4} \Bigg\} 
+ \cdots, \nonumber \\
S_{\textrm{het}} &=& \frac{1}{2\lambda_s^{8}} \int d^{10}x \sqrt{-g} e^{-\phi_d} \left[ R + \big( \nabla\phi_d \big)^2 - \frac{\tilde{\tilde{H}}_3^{2}}{2} - \frac{\textrm{Tr}\big(F_2^{2}\big)}{4} \right] + \cdots \nonumber,
\end{eqnarray}
respectively, where $\lambda_s$ is the strings length scale and $\phi_d$ denotes a dilaton field. $H_3$ is the field strength tensor of the NS-NS two-form $B_2$, while $\tilde{H}_3$ and $\tilde{\tilde{H}}_3$ stand for mixed contributions of the R-R two-form $A_2$ and the NS-NS two-form $B_2$, respectively, and the matrix-valued one-form $A_1$. $F_2$ is the field strength tensor of the R-R one-form $A_1$ and $\tilde{F}_4=dA_3+A_1\land H_3$ with $A_3$ being a three-form field. The dimensional reduction procedure gives effective actions for the considered theories, which can be collectively written as
\begin{eqnarray}
S_{\textrm{IIA/I/het}}^{(4)} = \frac{1}{16\pi} \int d^4x \sqrt{-g} \bigg\{ e^{-\phi_d} \left[ R + \big( \nabla\phi_d \big)^2 \right] - \chi F_2^{2} \bigg\} + \cdots,
\end{eqnarray}
where $\chi$ equals $1$ for the type IIA, $e^{-\frac{\phi_d}{2}}$ for the type I and $e^{-\phi_d}$ for the heterotic theory. The redefinition of the dilaton field $e^{-\phi_d}\to\Phi_d$ leads to the gravitational sector, which is proportional to the term $\Phi_d \Big[ R + \big( \nabla\Phi_d \big)^2 \Phi_d^{-2} \Big]$ for all the studied versions of the string theory. The two-form field becomes proportional to the term $\Phi_d^{\beta} F_2^{2}$ with $\beta$ equal to $0$ for the type IIA, $0.5$ for the type I and $1$ for the heterotic theory.

An interesting observation of the dilaton-matter coupling is that unless $\beta = 0$, there appears a direct coupling between the dilaton and matter fields. This allows a dynamical formation dilaton hair in the stationary limit. If such dilaton hair does not exist, a charged black hole has an inner apparent horizon, while if there is dilaton hair, there is no internal structure and only a space-like singularity appears \cite{Hansen:2014rua,Hansen:2015dxa}.

\subsubsection{Brans-Dicke theory}

It is also interesting to inspect the Brans-Dicke theory itself. As we have mentioned, one of the motivations is the dilaton field of the string theory. The low energy effective action of each string theory contains a sector with such a field
\begin{eqnarray}
S_{\textrm{d}} &=& \frac{1}{2\lambda_s^{D-1}} \int d^{D+1}x \sqrt{-g} e^{-\phi_d} \left[ R + \big( \nabla\phi_d \big)^2 \right],
\end{eqnarray}
where $D$ denotes a number of space dimensions. The field redefinition $\lambda_s^{1-D}\cdot e^{-\phi_d}\to\Phi_d\left(8\pi G_{D-1}\right)^{-1}$, where $G_{D-1}$ is the $D-1$-dimensional gravitational constant, gives the $\omega=-1$ limit of the Brans-Dicke theory.

The value of the Brans-Dicke parameter can be calculated in the weak field limit of the Randall-Sundrum braneworld model \cite{Randall:1999vf} according to $\omega = 1.5 \left( e^{\pm s/l}-1 \right)$, with $s$ being the distance between the branes and $l=\sqrt{-6\Lambda^{-1}}$ is the length scale of the anti-de Sitter space, while the sign in the exponent depends on the sign of the brane tension \cite{Garriga:1999yh}. The value of $\omega$ in these models is usually close to $-1.5$. When $\omega$ is less than $-1.5$, the kinetic term of the Brans-Dicke field in the Lagrangian is negative in the Einstein frame and hence the field acts as a ghost.

Based on the double-null simulations, we can easily observe that as $|\omega|$ approaches $1.5$, the Brans-Dicke field becomes more and more dynamic \cite{Hwang:2010aj}. In the ghost limit $\omega < -1.5$, the gravitational collapse can even induce an inflating space \cite{Hansen:2014rua,Hansen:2015dxa}.

\subsubsection{$f(R)$ gravity}

The gravitational sector of the action which corresponds to the scalar-tensor version of the $f(R)$ gravity is
\begin{eqnarray}
S_{f(R)} = \frac{1}{16\pi} \int d^4x\; \sqrt{-g}\; \Big[ f(\psi) + f^\prime(\psi) \left(R-\psi\right) \Big],
\end{eqnarray}
where $\psi$ is an auxiliary scalar field and $'$ denotes a derivation with respect to $\psi$. The field redefinition $f^\prime(\psi)\to\Phi_\psi$ allows to write the above action in the form
\begin{eqnarray}
S_{f(R)} = \frac{1}{16\pi} \int d^4x\; \sqrt{-g}\; \Big( \Phi_\psi R - V(\Phi_\psi) \Big),
\end{eqnarray}
which corresponds to a Brans-Dicke field with a potential when the coupling vanishes. Thus, the case of $\omega=0$ was interpreted as the $f(R)$ limit of the theory \cite{Hwang:2011kg,Chen:2015nma}. 

As was shown in \cite{Hwang:2011kg}, some forms of $f(R)$ may cause the gravitational collapse to induce a cusp singularity. This problem can be cured by adding an $R^{2}$ term, but still the $f(R)$ function is able to induce a bump of the Ricci scalar even outside the horizon \cite{Chen:2015nma}. In some aspects, this kind of models should be carefully ruled out and one needs to carefully choose model parameters to avoid this effect.

\begin{figure}
\begin{center}
\includegraphics[scale=0.27]{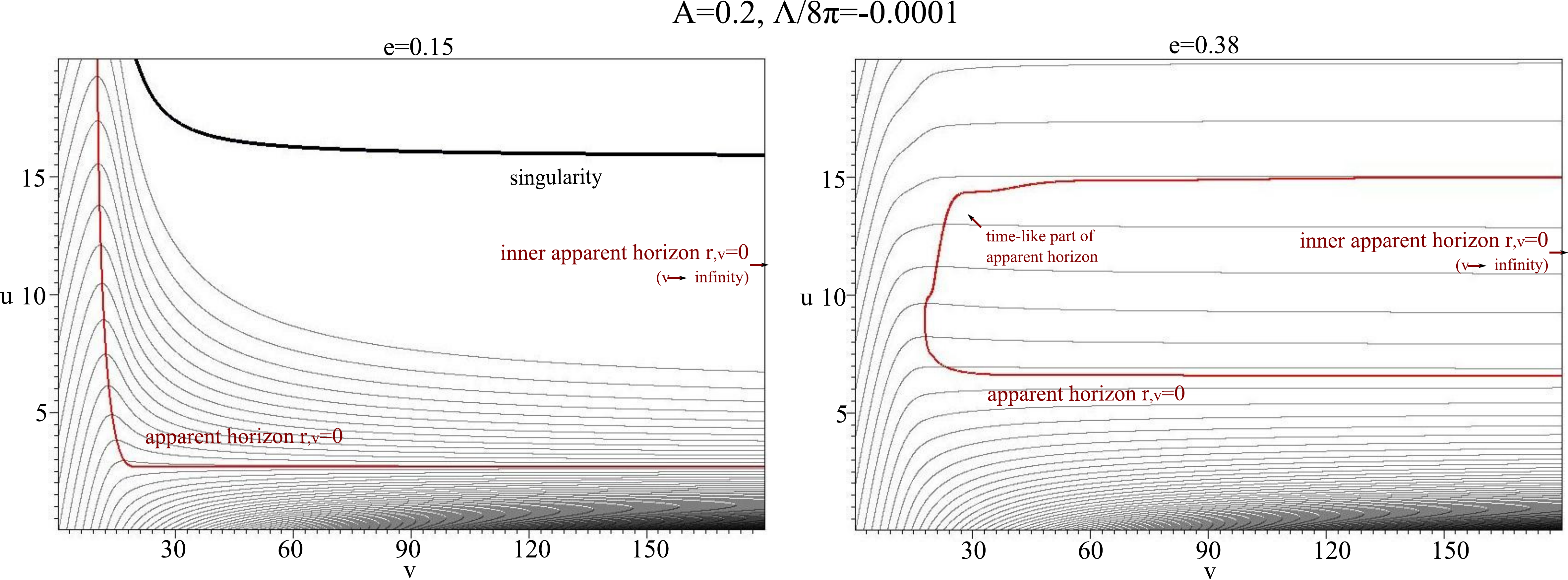}
\caption{\label{fig:P=0}Various causal structures of a charged black hole in three-dimensions, by varying $e$ and fixing $A = 0.2$ and $\Lambda/8\pi = -0.0001$.}
\includegraphics[scale=0.25]{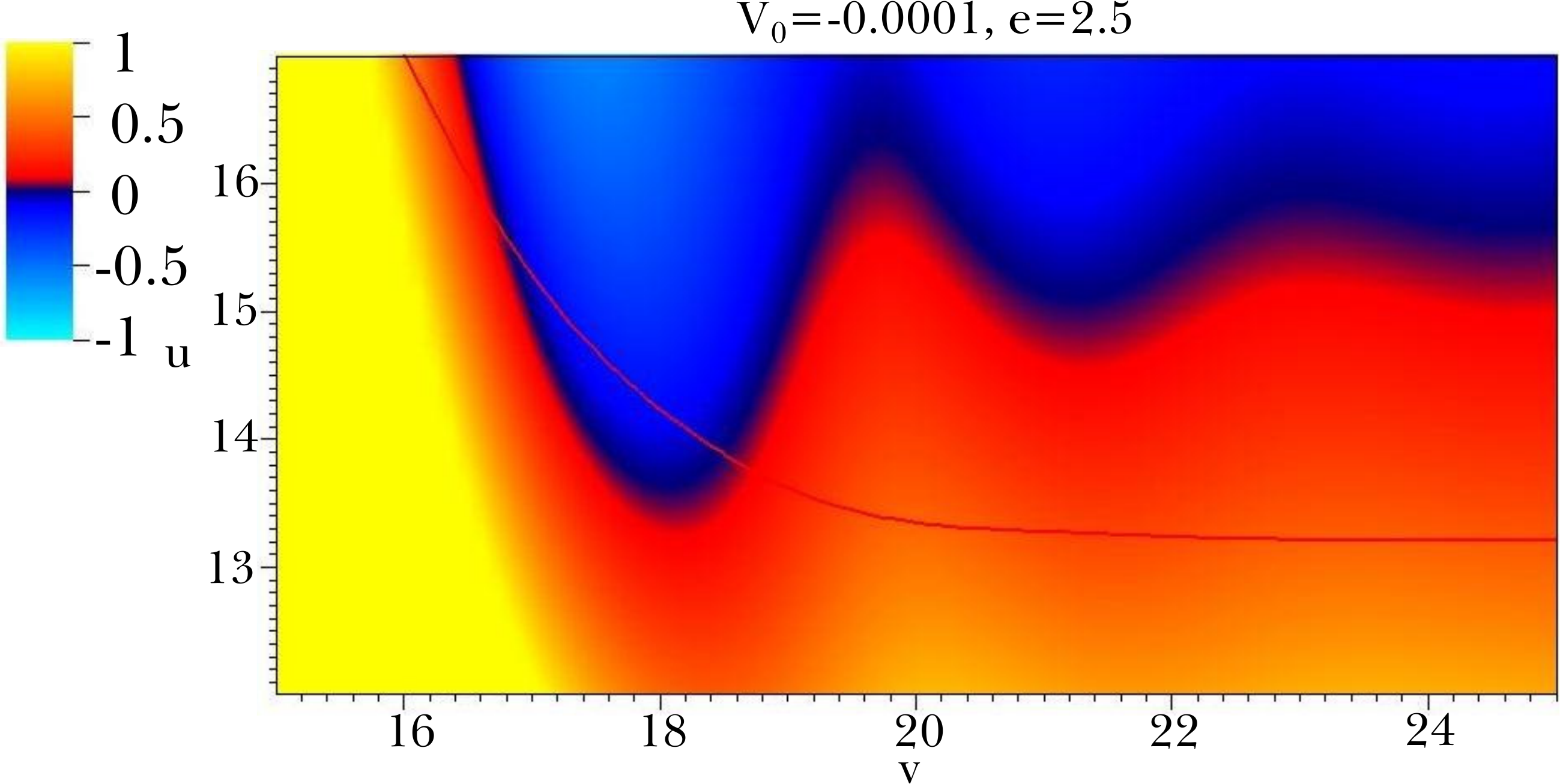}
\includegraphics[scale=0.17]{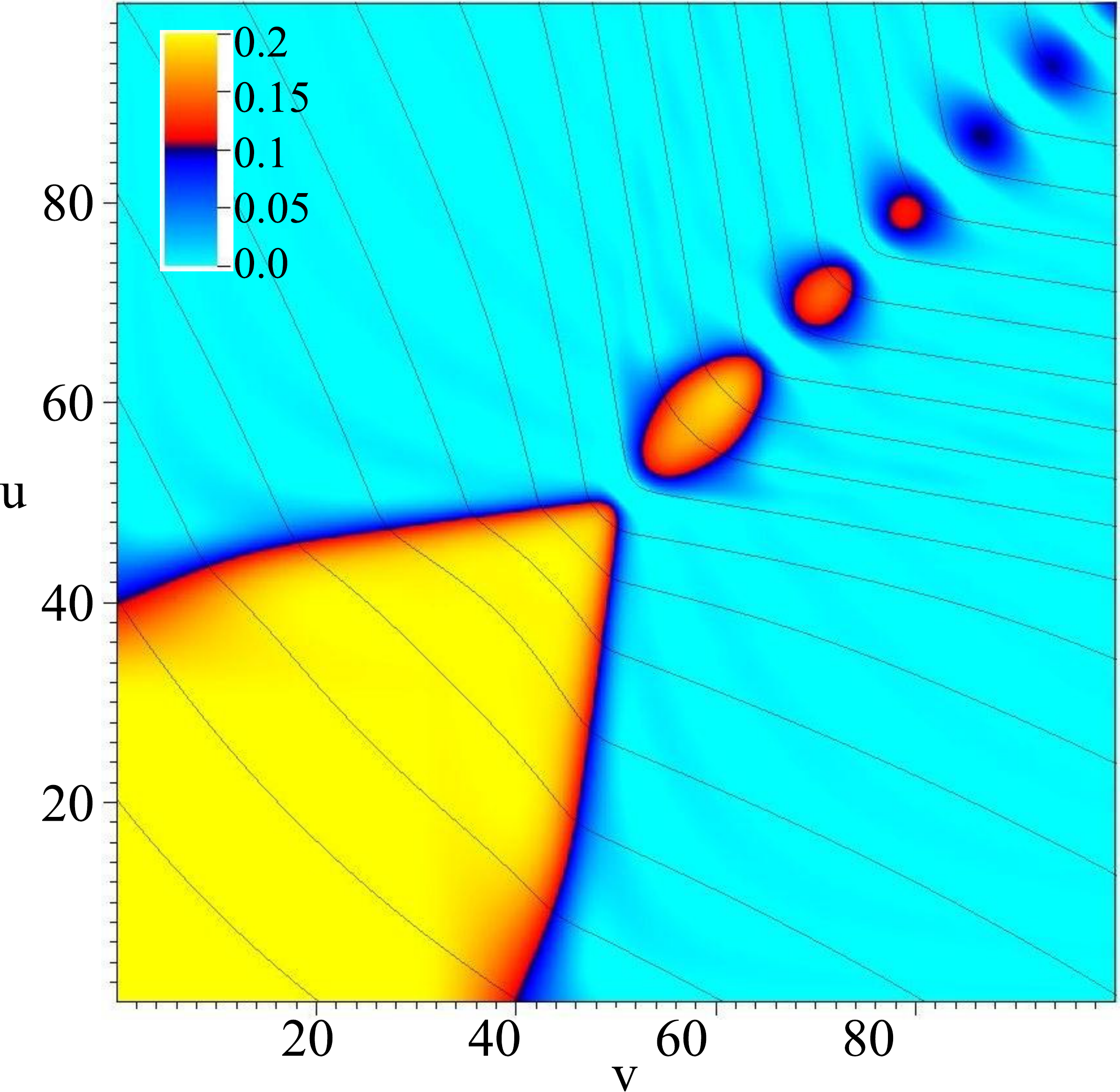}
\caption{\label{fig:L0001_Q_near}Left: Behavior of electric charge near a black brane. Right: An example of a bubble collision with hyperbolic symmetry, where color denotes a field value (yellow  and blue are false and true vacuum regions, respectively).}
\end{center}
\end{figure}

\subsection{Dimensions}

Another way of extending the double-null formalism is to change the number of dimensions, for example, choose $ds^{2} = -\alpha^{2}(u,v) du dv + r^{2}(u,v) d\Omega_{D-2}^{2}$, where $d\Omega_{D-2}$ denotes a $(D-2)$-sphere. In addition, it is theoretically interesting to consider a lower dimension, where the only suitable lower dimension is three: $ds^{2} = -\alpha^{2}(u,v) du dv + r^{2}(u,v) d\varphi^{2}$, assuming the circular symmetry. Then the energy-momentum tensor is similar to the four-dimensional, but the Einstein tensor is different, for example, for three dimensions \cite{Hwang:2011mn}
\begin{equation}
G_{uu}^{(3)} = -\frac{1}{r} \left(f_{,u}-2fh \right),\;\;\;
G_{uv}^{(3)} = \frac{f_{,v}}{r},\;\;\;
G_{vv}^{(3)} = -\frac{1}{r} \left(g_{,v}-2gd \right),\;\;\;
G_{\varphi\varphi}^{(3)} = -4\frac{r^{2}}{\alpha^{2}} d_{,u}.
\end{equation}
This new aspect of the Einstein tensor is the origin of new features. One can consider lots of different settings of higher dimensions. Fig. \ref{fig:P=0} reports several causal structures of charged black holes in an anti-de Sitter background \cite{Hwang:2011mn}.

\subsection{Topologies}

One may also change the topology of the spatial part:
\begin{eqnarray}
d\Omega_{\kappa=1}^{2} &=& d\theta^{2} + \sin^{2} \theta d\varphi^{2},\\
d\Omega_{\kappa=0}^{2} &=& dx^{2} + dy^{2},\\
d\Omega_{\kappa=-1}^{2} &=& d\chi^{2} + \sinh^{2} \chi d\varphi^{2},
\end{eqnarray}
where the symmetry is spherical ($\kappa = 1$), planar ($\kappa = 0$), and hyperbolic ($\kappa = -1$), respectively. The variables cover the ranges $0 \leq \varphi \leq 2\pi$, $0 \leq \theta \leq \pi$, $-\infty \leq x,y \leq \infty$, and $0 \leq \chi \leq \infty$. Static solutions for each symmetry are known as black holes, black branes, and topological black holes, respectively. Because of the choice of $\kappa$, the only difference is in the $uv$-component of the Einstein tensor:
\begin{eqnarray}
G_{uv} = \frac{1}{2r^{2}} \left( 4 rf_{,v} + \kappa \alpha^{2} + 4fg \right).
\end{eqnarray}
Due to this difference, one may see different phenomenology of various black objects. Especially, the mass function becomes
\begin{eqnarray}
m(u,v) = \frac{r}{2} \left( \kappa + \frac{4 r_{,u} r_{,v}}{\alpha^{2}} + \frac{Q^{2}}{r^{2}} - \frac{\Lambda}{3} r^{2} \right).
\end{eqnarray}

\subsubsection{$\kappa = 0$: black branes}

For $\kappa = 0$, at the $u = v = 0$ point, one can give $A_{u} = q = 0$. At this moment, $\alpha(0,0) = \sqrt{4 f(0,0) g(0,0)/\mathcal{D}}$ with $\mathcal{D} = 2m(0,0)/r_{0} + 8\pi V_{0} r_{0}^{2}/3$ with the corresponding vacuum energy $V_{0}$. Hence, one can have a gravitational collapsing situation ($f < 0$ and $g > 0$) only if $\mathcal{D} < 0$, and hence in anti de Sitter space $V_{0} < 0$ (if $m(0,0) \geq 0$). Hence, one can see dynamical responses of observables near the horizon and there can be interesting issues in the context of the AdS/CFT correspondence (the left panel of Fig. \ref{fig:L0001_Q_near}) \cite{Hansen:2013vha}.

\subsubsection{$\kappa = -1$: bubble collisions}

If $\kappa = 0$ or $-1$, it is physically possible to choose $r_{,u}$ and $r_{,v}$ both positive. Then this is no more a gravitational collapse, but describes a domain wall or bubble collisions. As Coleman-DeLuccia vacuum bubbles \cite{Coleman:1980aw} are nucleated, each bubble has an $O(3,1)$ symmetry. If there are two bubbles, then one can choose a good coordinate such that the symmetry is reduced by $O(2,1)$ which is described by a hyperbolic symmetry ($\kappa = -1$) \cite{Hawking:1982ga}. If the bubble wall is large enough, then the approximation of $\kappa = 0$ is also sound. One can see dynamical behaviors of bubble collisions which can be realized by cosmological contexts (the left panel of Fig. \ref{fig:L0001_Q_near}) \cite{Hwang:2012pj}.

\section{\label{sec:app}Applications}

In this section, we summarize applications of the double-null formalism, especially issues that can only be understood by numerical simulations, where this covers a broad range of topics from quantum gravity to astrophysics and cosmology.

\subsection{Black hole physics}

\subsubsection{Applications to (loop) quantum gravity}

There is a problem in transferring a notion of time between classical and quantum formulations of gravity. One of the existing proposals of quantifying time in dynamical quantum gravitational systems is using the evolving matter as an intrinsic `clock'. The 2+2 formalism was used to check whether a scalar field may serve as a time variable during a dynamical evolution of a sole scalar field and also of coupled multi-component matter-geometry systems, which were an electrically charged scalar field and a scalar field evolving within the Brans-Dicke gravity. The focus was mainly on the region of high curvature neighboring the emerging singularity, which is essential for the quantum gravity applications. A successful description of the passage of time in a gravitational system requires the following two conditions to be fulfilled during a selected part of an evolution: (i) the selected spacetime slices, parametrized by a time variable, ought to be spacelike and (ii) the chosen time parametrization should remain monotonic in the region of interest.

\begin{figure}
\begin{center}
\includegraphics[scale=0.145]{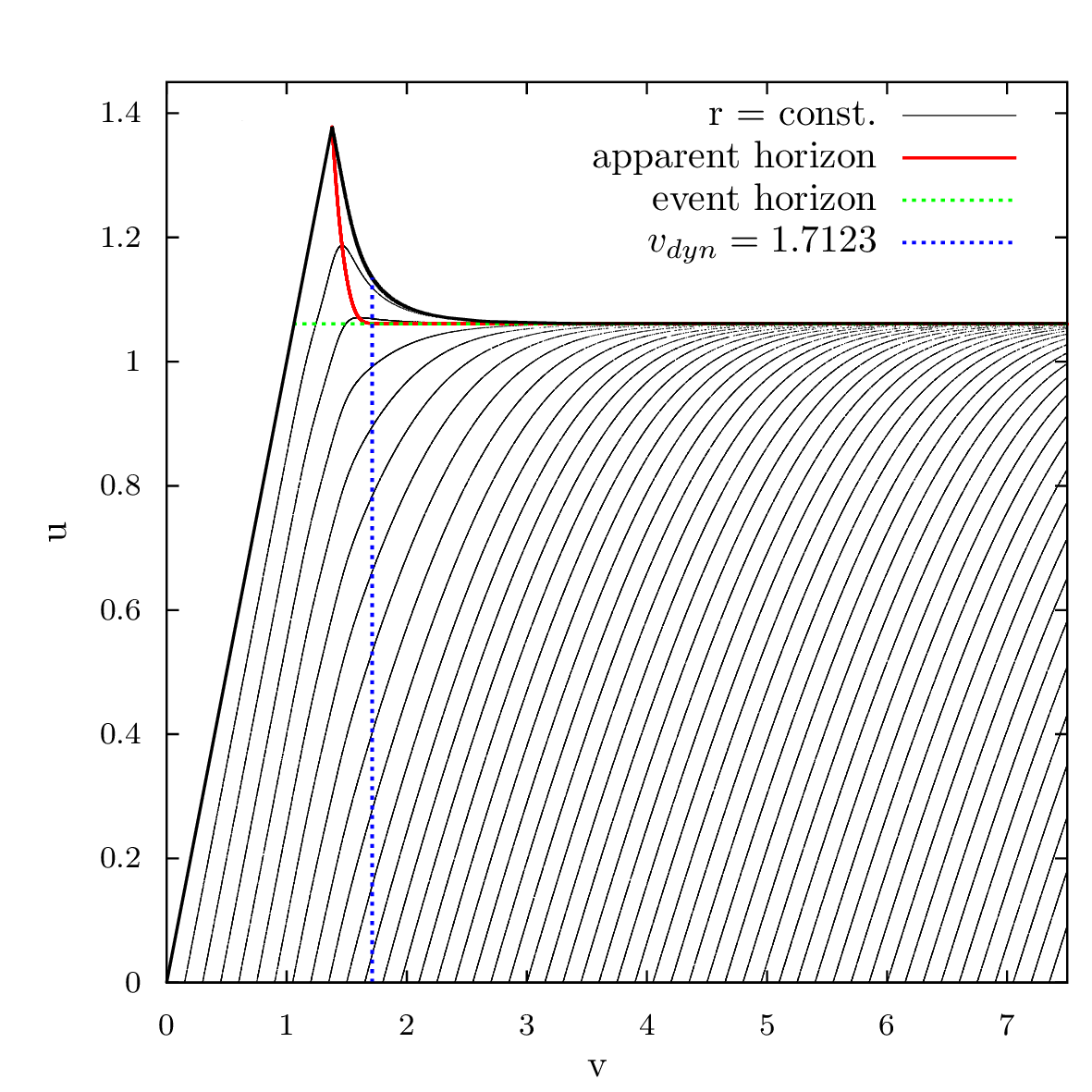} \includegraphics[scale=0.142]{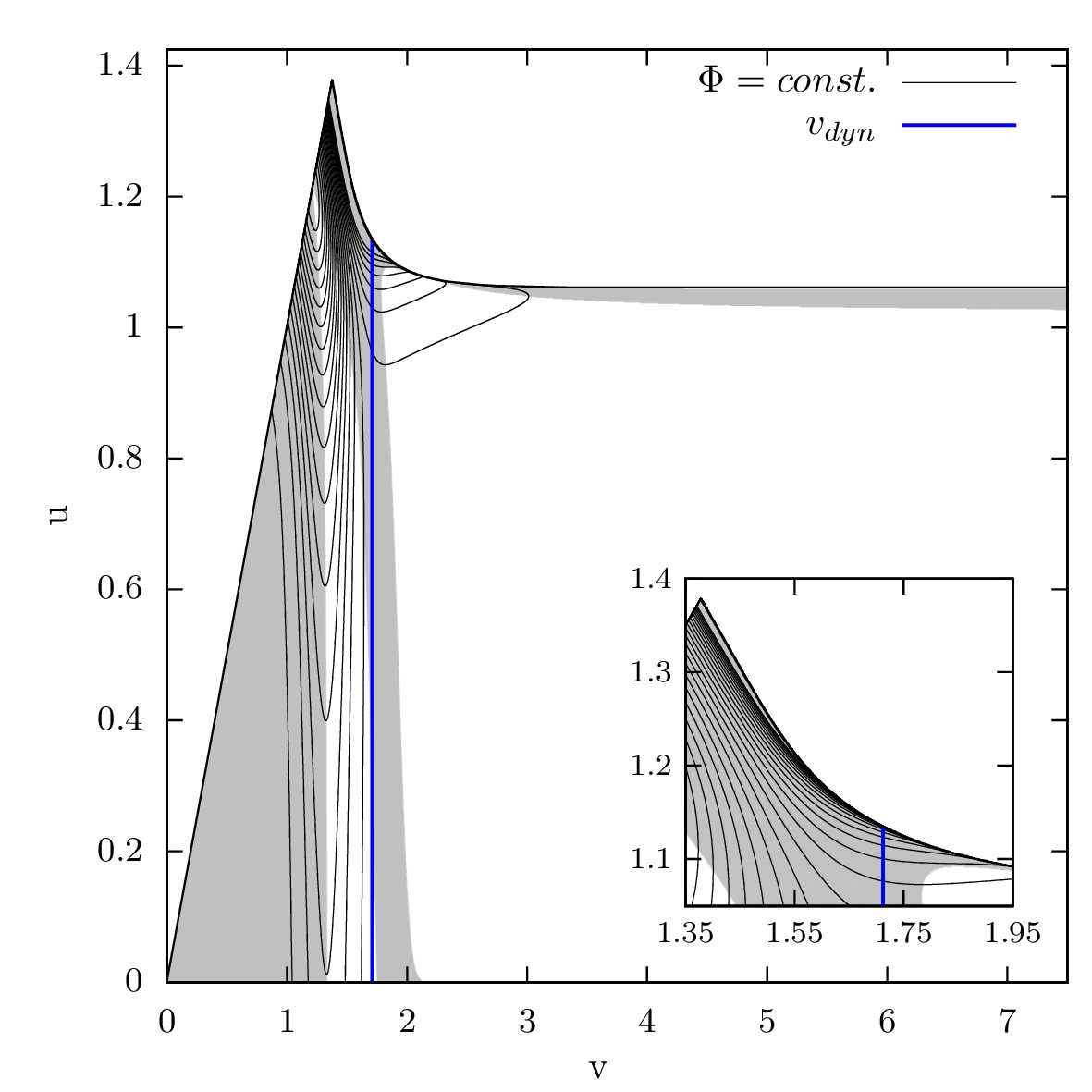}
\caption{\label{fig:mt}Left: Penrose diagram of a dynamical Schwarzschild spacetime formed during the scalar field gravitational
collapse in double null formalism. The null hypersurface $v_{dyn}$ denotes a border between dynamical and non-dynamical spacetime regions. Right: The scalar field constancy lines within the developed spacetime. Gray areas indicate spacetime regions, in which the hypersurfaces are spacelike. The dynamical region neighboring the central singularity was magnified.}
\end{center}
\end{figure}

An example of the outcomes of the investigated gravitational evolutions is presented in Fig. \ref{fig:mt}\cite{prd92(2015)064031}. Overall, using scalar fields as time variables within whole spacetimes during dynamical gravitational evolutions of coupled matter-geometry systems is limited. First, the two above conditions which are necessary for treating the field as a time measurer are not fulfilled in the whole spacetimes. Second, the vicinity of Cauchy horizons should be excluded from the analyses. Third, the forms of the field evolution equations should be checked for various values of parameters which they contain, because the possibility of using the specific scalar field as a time variable may be excluded in some cases (e.g., when the equation of motion of the field reduces to the wave equation due to a specific choice of its parameters). Fortunately, only the last of the above difficulties applies to a close proximity of the singularity emerging in the spacetime and this region of high curvature is of crucial importance for the gravity quantization.

\begin{figure}
\begin{center}
\includegraphics[scale=0.27]{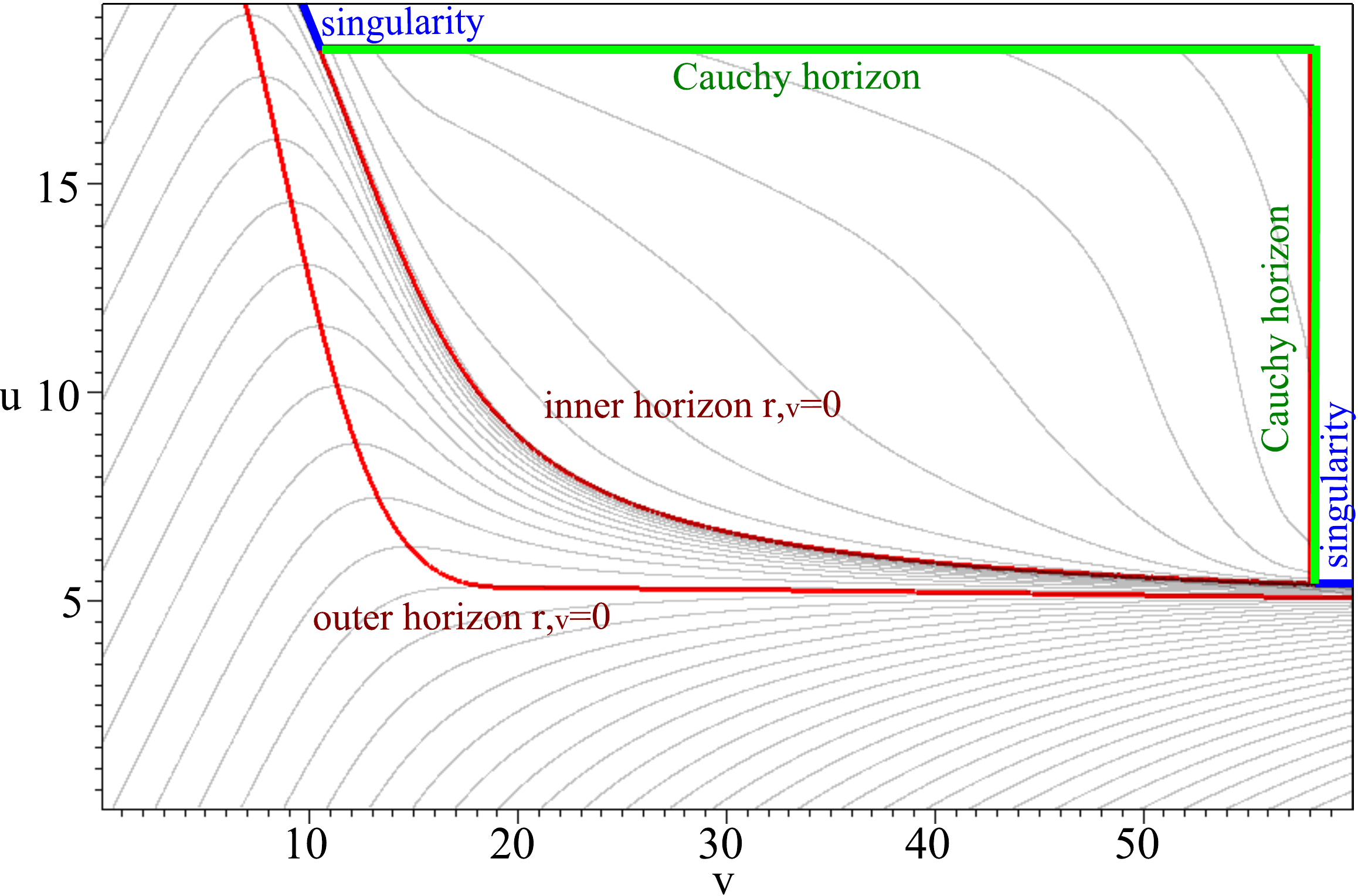}
\includegraphics[scale=0.3]{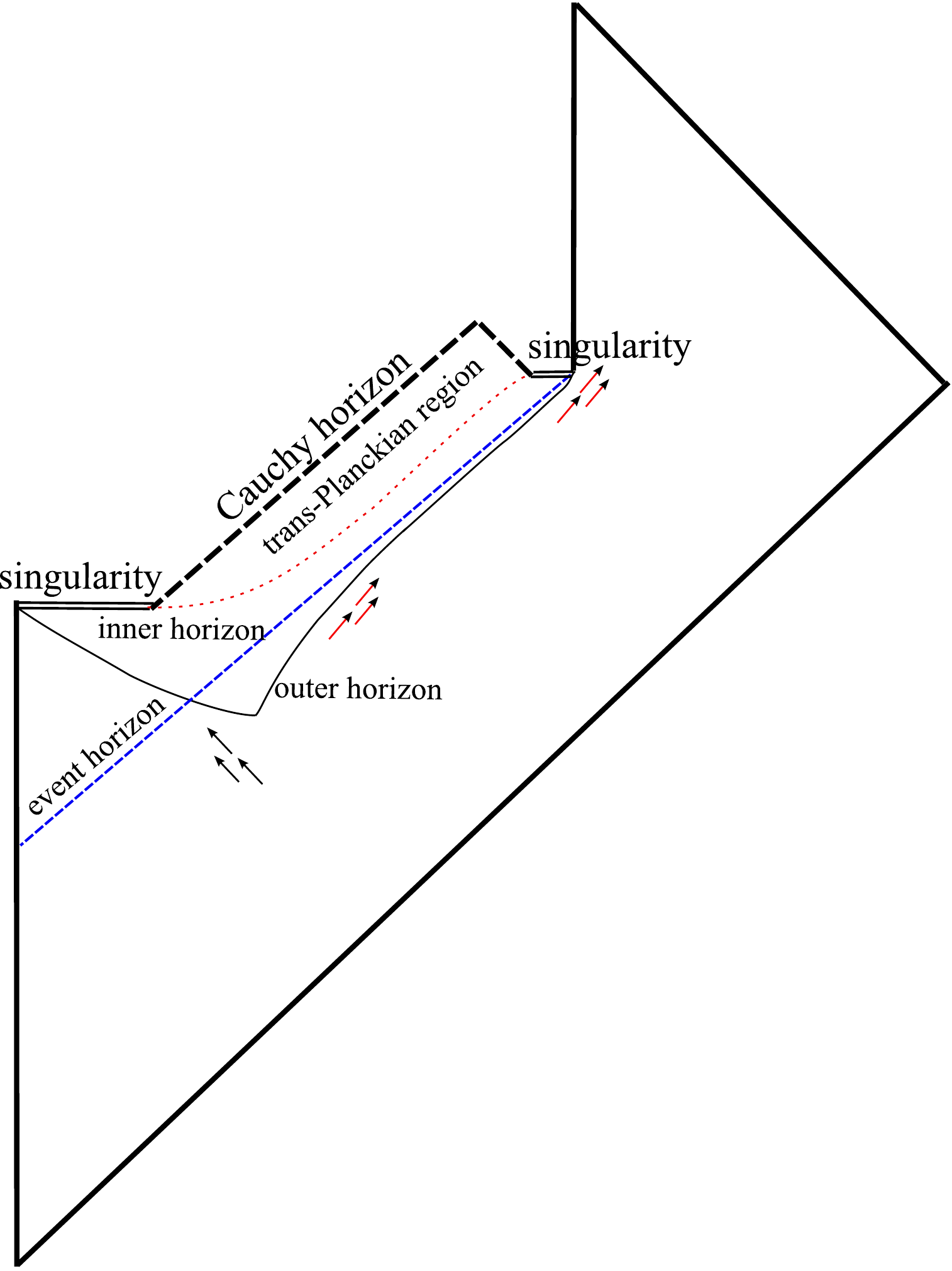}
\caption{\label{fig:charged_neutral_contour}Numerical simulations (left) and causal structures (right) of dynamical charged black holes, including gravitational collapse, Hawking radiation, and discharging effects.}
\end{center}
\end{figure}

\subsubsection{Cosmic censorship and internal structure of charged black holes}

The cosmic censorship conjecture is related to the deterministic nature of general relativity. According to the strong form of the conjecture, no observer can see effects from a singularity unless the observer hits a singularity (except for the initial singularity of the universe). In its weak version, singularities cannot be naked and must be hidden by the event horizon (again, except for the initial singularity of the universe); but this opens a possibility that one can see effects of the singularity inside the event horizon.

For the static charged black hole solution, there is an inner Cauchy horizon where one can see effects of a time-like singularity as one crosses it. If charge $Q$ is larger than mass $M$, then the time-like singularity can be even outside the horizon and be naked. Then can weak and strong cosmic censorship conjectures be violated in charged black holes?

The important observation is that these conjectures are related to the dynamical behaviors of fields. Regarding the weak version of the cosmic censorship, one can prepare asymptotic charge and mass that satisfies $Q > M$. However, as one derives the gravitational collapse of the combination, it is impossible to form a $Q > M$ combination at the horizon \cite{Hwang:2010im}. This is due to the repulsive interactions of the field. This preserves the weak cosmic censorship.

Regarding the strong cosmic censorship, theoretically, it was observed that the inner horizon is unstable via an infinite blue-shift \cite{Simpson:1973ua,Brady:1995ni}. However, if it is unstable, then we cannot trust the static solution and we essentially need numerical computations \cite{Hod:1998gy}. Then we can see mass inflation \cite{Poisson:1990eh}, where the mass function increases exponentially as one approaches the inner Cauchy horizon (the left panel of Fig. \ref{fig:bubble}) \cite{Hong:2008mw}.

If one adds Hawking radiation, then even this causal structure is modified \cite{Sorkin:2001hf}. By mimicking discharging effects by Schwinger processes, one can draw the realistic causal structure of dynamical charged black holes (Fig. \ref{fig:charged_neutral_contour}) \cite{Hong:2008mw}.

\begin{figure}
\begin{center}
\includegraphics[scale=0.25]{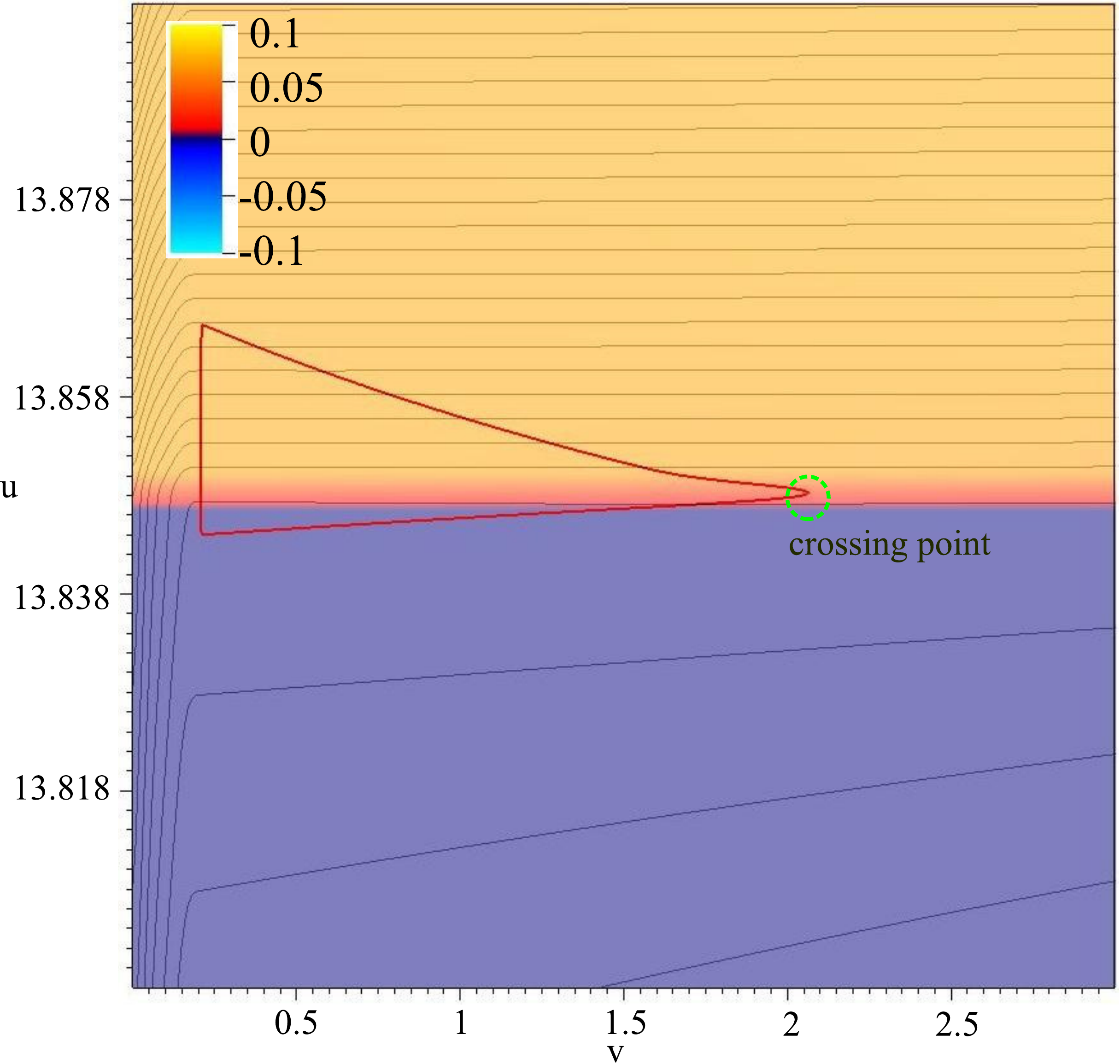}
\includegraphics[scale=0.65]{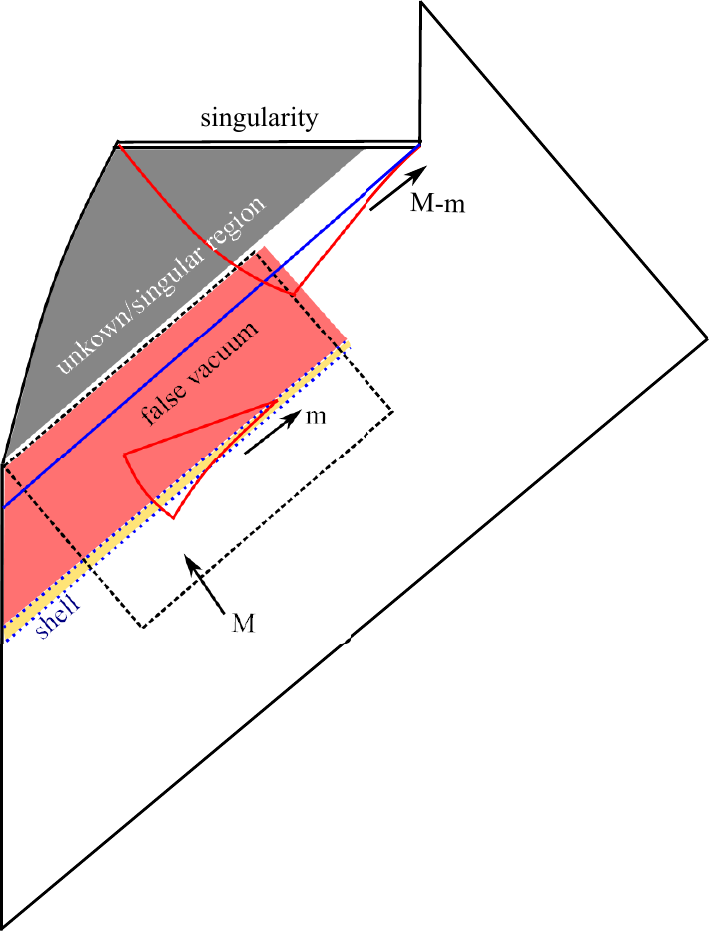}
\caption{\label{fig:crossing}Numerical simulations (left) and causal structures (right) of a black hole with a false vacuum core, where the yellow region is false and the blue region is true vacuum.}
\includegraphics[scale=0.25]{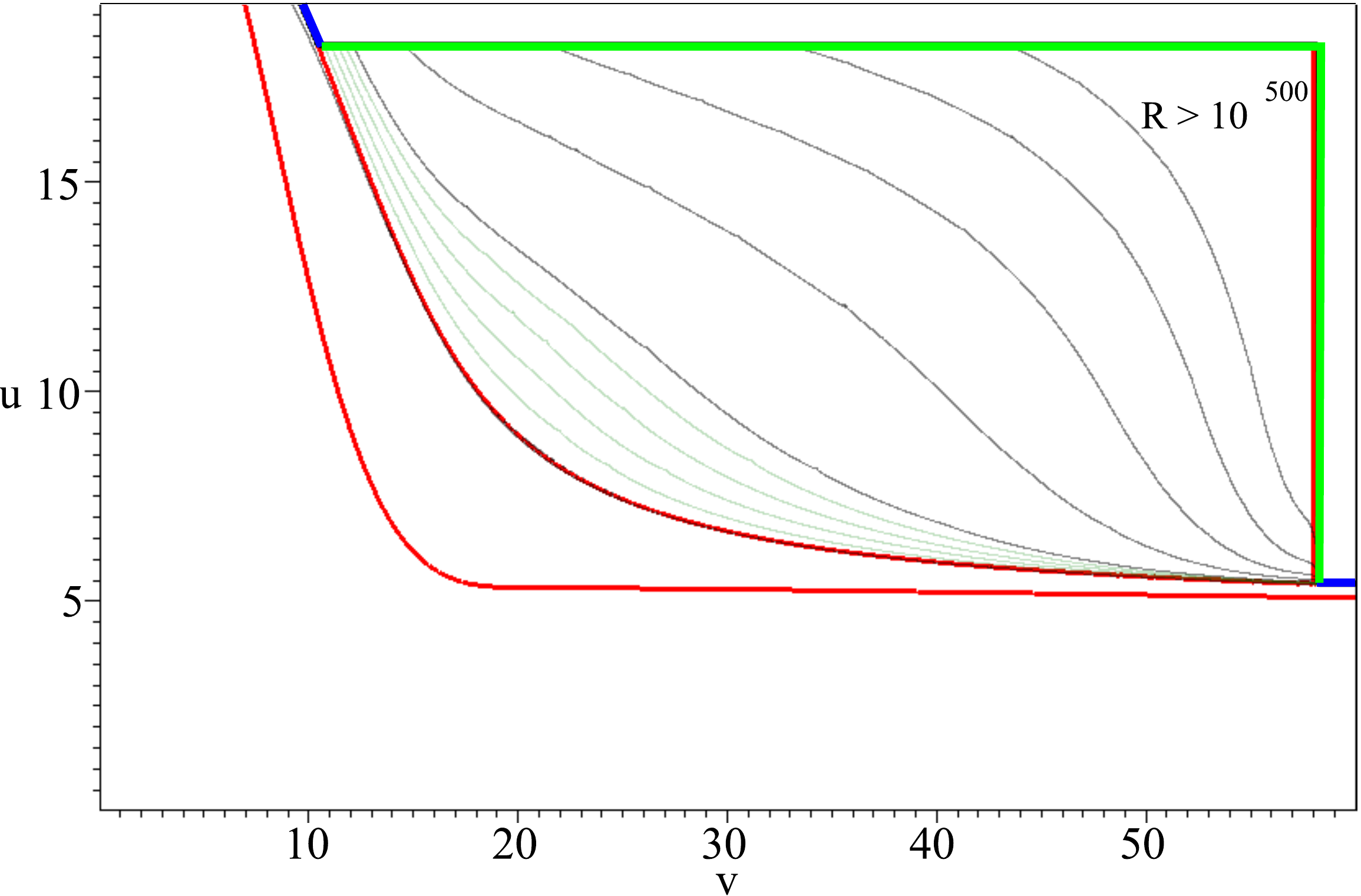}
\includegraphics[scale=0.5]{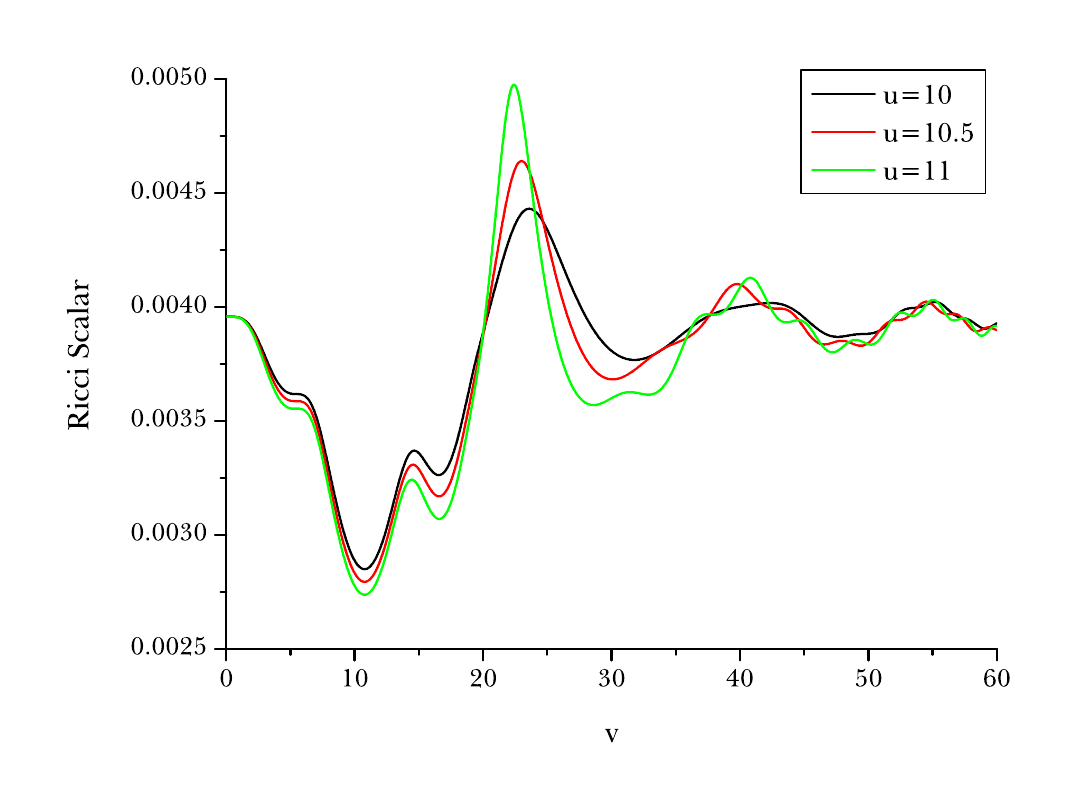}
\caption{\label{fig:Ricci}Left: the Ricci scalar for inside the black hole, where the contours denote $\log |R| =
0, 20, 40, 60, 80, 100, 200, 300, 400, 500$. Right: example behaviors of the Ricci scalar of the Starobinsky model \cite{Starobinsky:2007hu}.}
\end{center}
\end{figure}

\subsubsection{Resolving and probing singularity}

Understanding and resolving a singularity inside a black hole is a very important task of quantum gravity. There may be several approaches, either we modify gravity which is free from singularities or we find a solution that is potentially free from singularity. There have been several models to mimic this, so-called regular black hole models. The simplest model introduces a false vacuum core inside the black hole \cite{Frolov:1988vj}. Based on Vaidya metric approximation, one can see that there is no formation of a singularity due to the repulsive force of the false vacuum. However, the genuine physics is dynamical and we need real simulations to confirm whether there is indeed no singularity or not.

The result is depicted in Fig. \ref{fig:crossing} \cite{Hwang:2012nn}. Originally, a regular black hole model has no singularity due to the violation of global hyperbolicity \cite{Chen:2014jwq}. However, in simulations, it should be satisfied. Then the only mechanism to avoid a singularity is the violation of the energy condition due to Hawking radiation. This can be demonstrated by numerical simulations, by tuning initial conditions (Fig. \ref{fig:crossing}) \cite{Hwang:2012nn}. The observation is that a circular shaped apparent horizon is possible \cite{Hayward:2005gi}, at least instantly. However, the negative energy of the Hawking radiation is not enough to resolve singularities of the entire causal structure (the right panel of Fig. \ref{fig:crossing}) that is consistent with previously expected diagrams (we can call it a semi-regular black hole) \cite{Yeom:2008qw}.

Another important issue is a singularity near the inner horizon. Due to the mass inflation, curvature quantities should increase exponentially (the left panel of Fig. \ref{fig:Ricci}) \cite{Hong:2008mw}. \footnote{Some authors investigated the phenomenon of a shock-wave like singularity formation \cite{Eilon:2016osg}, the so-called Marolf-Ori singularity \cite{Marolf:2011dj}, though its existence in the realistic gravitational collapse is less clear.} There is no consensus to resolve this problem, but modified gravity can give a hint. For example, in $f(R)$ gravity, there is mass inflation but the Ricci scalar $R$ can be bounded (the right panel of Fig. \ref{fig:Ricci}) \cite{Hwang:2011kg}.

\begin{figure}
\begin{center}
\includegraphics[scale=0.6]{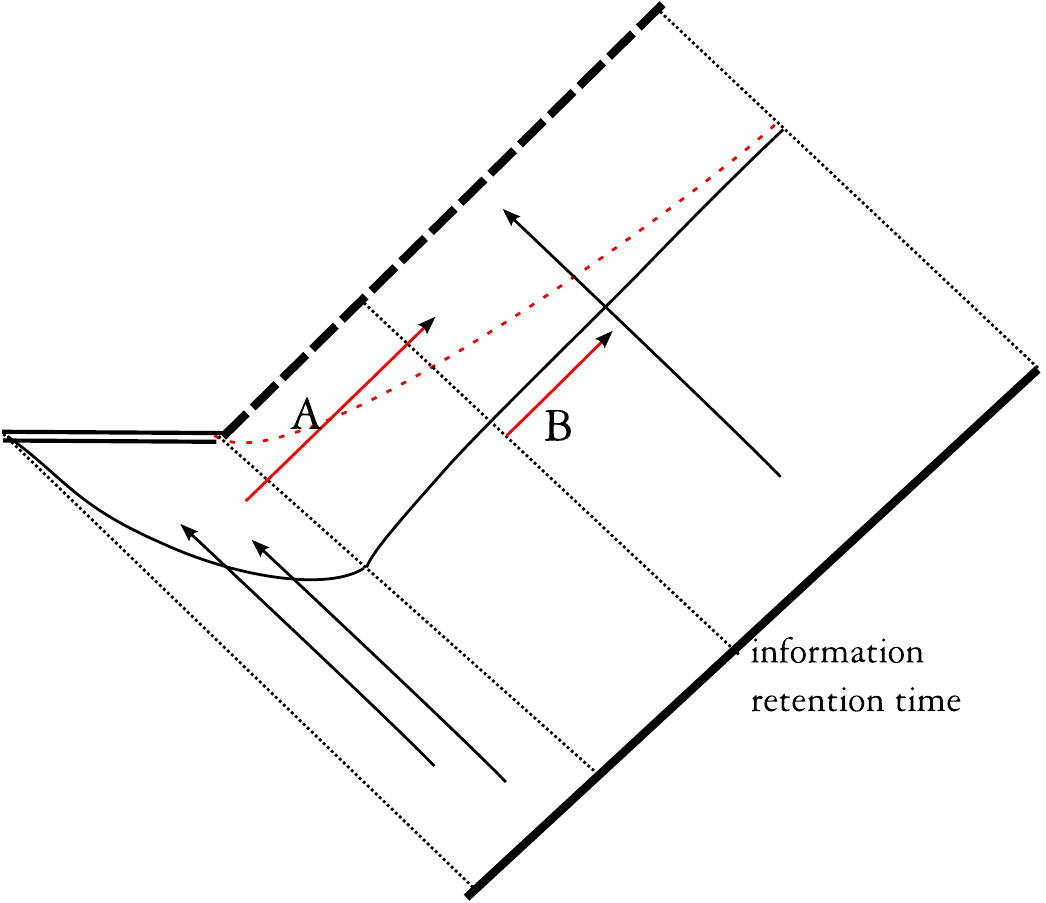}
\includegraphics[scale=0.7]{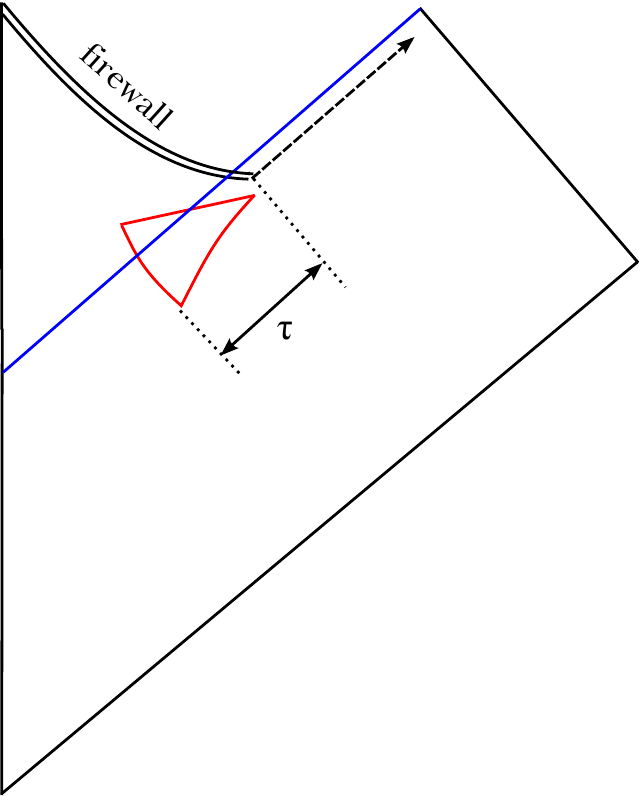}
\caption{\label{fig:complementarity}Left: duplicated information can be observed for a charged black hole, if we assume a large number of scalar fields that contribute to evaporation. Right: if a firewall grows around the apparent horizon, it can be eventually naked, where $\tau$ is the information retention time.}
\end{center}
\end{figure}

\subsubsection{Information loss problem of black holes and related topics}

The information loss problem is related to the tension between quantum mechanics and semi-classical gravity \cite{Hawking:1976ra}. Many proposals that resolve the information loss paradox are based on semi-classical pictures, including black hole complementarity \cite{Susskind:1993if}, the firewall proposal \cite{Almheiri:2012rt}, or the regular black hole picture \cite{Hayward:2005gi}. We summarize several important applications:
\begin{itemize}
\item[--] Based on our semi-classical simulations of charged black holes \cite{Hong:2008mw} and regular black holes \cite{Yeom:2008qw}, one can show that dynamical causal structure inside the black hole can give counterexamples of black hole complementarity (the left panel of Fig. \ref{fig:complementarity}), especially in the large $N$ limit \cite{Yeom:2009zp}.
\item[--] Based on the causal structure of a semi-regular black hole, it is reasonable to guess that the firewall should be naked if it needs to grow up to the apparent horizon (the right panel of Fig. \ref{fig:complementarity}) \cite{Hwang:2012nn}. This conclusion was also confirmed by rather conservative arguments \cite{Chen:2015gux}.
\item[--] It is fair to say that a false vacuum core hardly realizes a totally regular spacetime (the right panel of Fig. \ref{fig:crossing}) \cite{Hwang:2012nn}. Hence, the regular black hole picture is still a limited and heuristic picture that has to rely on yet unknown physics.
\end{itemize}

In addition, by considering vacuum bubbles and semi-classical effects, one can numerically obtain a negative-energy radiating bubbles \cite{Hwang:2010gc}. If this negative energy is condensed, it can generate a bubble universe inside a black hole \cite{Hansen:2009kn}. Then this strongly supports, at least, the effective loss of information \cite{Sasaki:2014spa}.

\begin{figure}
\begin{center}
\includegraphics[scale=0.5]{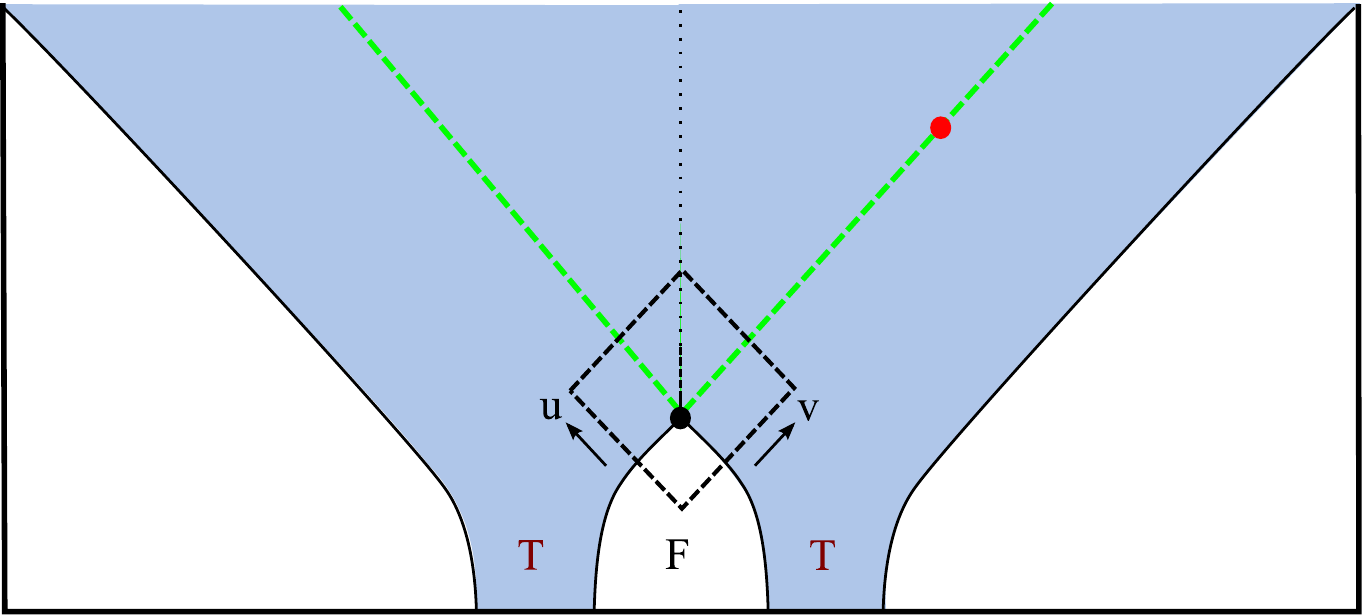}
\includegraphics[scale=0.15]{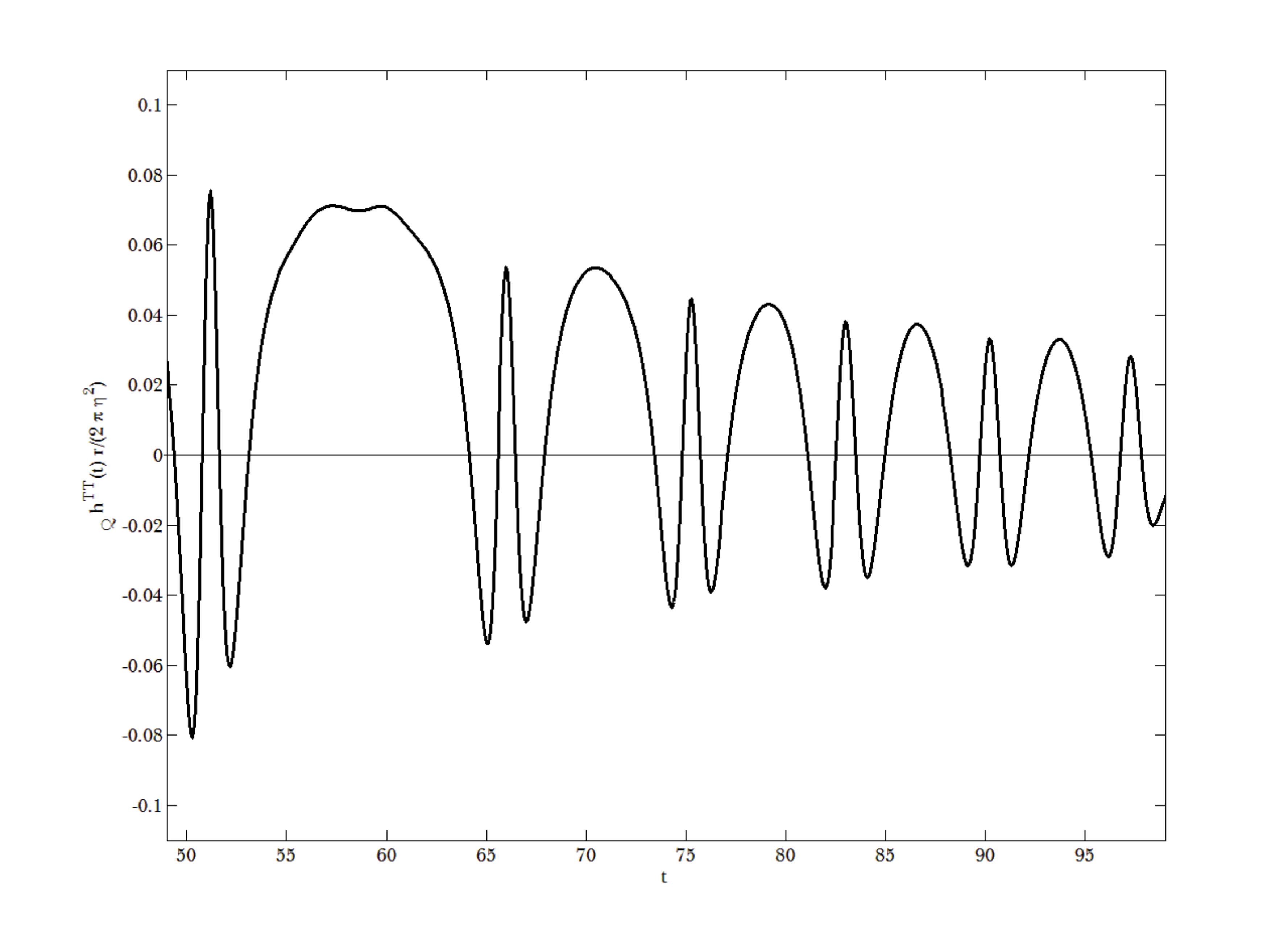}
\caption{\label{fig:figure}Left: the physical situation (red dot) when we observe bubble collisions via gravitational waves. Right: an example of $_{\mathrm{Q}}h_{ij}^{\mathrm{TT}}$ for the right of Fig. \ref{fig:L0001_Q_near}.}
\end{center}
\end{figure}

\subsection{Cosmology and particle astrophysics}

\subsubsection{Bubble dynamics and applications for gravitational waves}

As we previously mentioned, by giving the planar or hyperbolic symmetry, one can describe dynamical bubbles \cite{Hwang:2012pj}. If such a bubble collision happens in a certain place of our universe, its effect can be transferred to a local observer, e.g., by gravitational waves. Based on the quadrupole approximation, one can present the amplitude of a gravitational wave such that \cite{Kim:2014ara}
\begin{equation}
_{\mathrm{Q}}h_{ij}^{\mathrm{TT}}\left( t,\mathbf{x}\right) =\frac{4}{r}%
\Lambda _{ij,kl}\left( \mathbf{n}\right) I_{kl}\left( t_{\mathrm{R}}\right)
\,,
\end{equation}
where
\begin{equation}
I_{kl}\left( t_{\mathrm{R}}\right) \equiv \int d^{3}x^{\prime }T^{kl}\left(
t_{\mathrm{R}},\mathbf{x}^{\prime }\right) \,.
\end{equation}%
Hence, one can regard the amplitude of a gravitational wave as proportional to the integration of the energy-momentum tensor. After some approximations, one can obtain the behavior such as the one shown in Fig. \ref{fig:figure} \cite{Kim:2014ara}.

\subsubsection{Strong gravitational back-reactions based on dark matter models}

From among a large variety of dark matter models, the gravitational collapse in double-null formalism was investigated within a theory consisting of a complex scalar field $\chi$ with a quartic self-interaction, charged under an Abelian gauge field $P_\beta$. A coupling of the field to the standard matter sector represented by an electrically charged scalar field was realized via a kinetic mixing between the introduced gauge and electromagnetic fields \cite{jhep11(2015)012}. The adequate Lagrangian is
\begin{equation}
\mathcal{L}_{\mathrm{DM}} = - \left(\nabla_\beta + i\tilde{e} P_\beta\right)\chi \left(\nabla^\beta - i\tilde{e} P^\beta\right)   \chi^\ast
- \frac{1}{4} B_{\beta \sigma} B^{\beta \sigma}
- \frac{\alpha_{DM}}{4} B_{\beta \sigma} F^{\beta \sigma} - V\left(|\chi|^2\right),
\end{equation}
where $\tilde{e}$ is a coupling constant between $P_\beta$ and $\chi$, $B_{\beta \sigma}\equiv\partial_\beta P_\sigma-\partial_\sigma P_\beta$. The potential is given by $V\left(|\chi|^2\right) = \frac{m^2}{2}|\chi|^2 + \frac{\lambda_{DM}}{4}|\chi|^4$. $m^2$ and $\lambda_{DM}$ are a square of the scalar field mass parameter and its quartic self-interaction coupling constant, respectively.

The model may describe one or two dark matter candidates. If the vacuum expectation value (vev) is non-zero, the dark matter candidate is the massive gauge boson, often called $Z^\prime$ or dark photon. When the scalar does not possess the vev, the candidate can be the scalar or the gauge boson.

The type of a black hole emerging from the dark matter collapse depends on the value of the mass parameter squared. The non-zero vev of the complex scalar field favors the formation of dynamical Schwarzschild black holes below some critical value of $m^2$, while for the vanishing vev the Reissner-Nordstr\"{o}m black holes form. The spacetime structures are in these cases analogous to the ones shown in Fig. \ref{fig:neutral} and on the left panel of Fig. \ref{fig:bubble}, respectively.

\begin{figure}
\begin{center}
\includegraphics[scale=0.25]{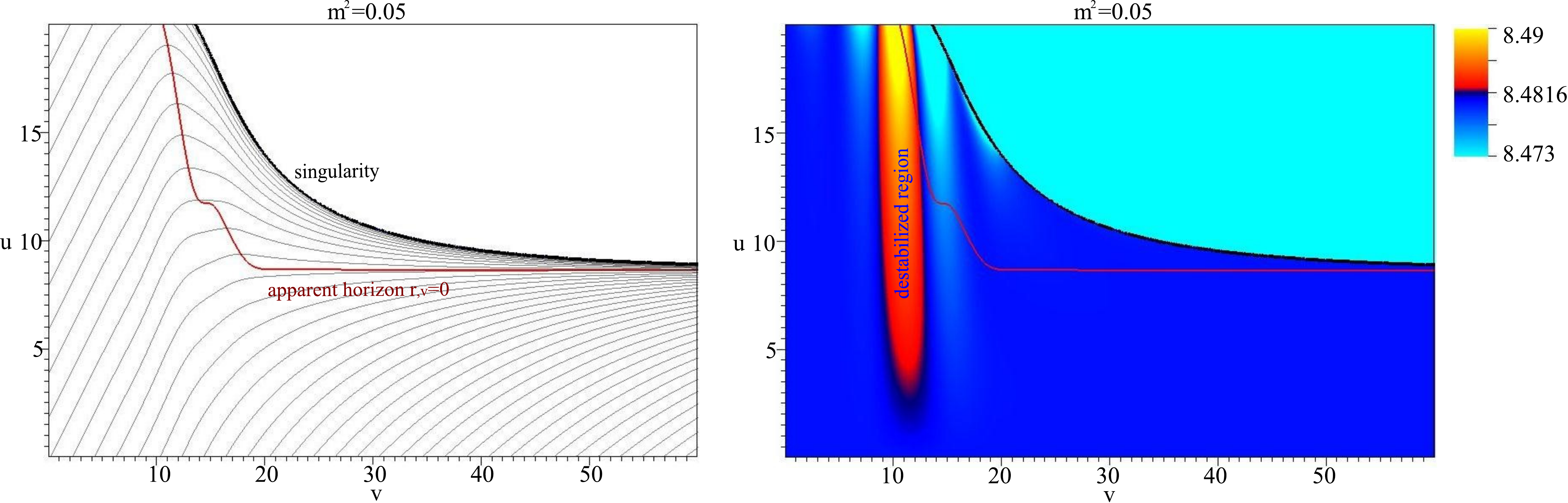}
\caption{\label{fig:plots}An example of destabilized moduli fields via the gravitational collapse.}
\end{center}
\end{figure}

\subsubsection{Strong gravitational back-reactions based on string-inspired models}

The double-null simulation can be applied to strong gravitational phenomena of string-inspired models. One interesting example is about the string compactification models. If extra dimensions are compactified, then it generates various moduli fields, which can couple to various matter sectors \cite{Conlon:2010jq}. Hence, the gravitational collapse procedure can give back-reactions to the moduli field that can be observed by asymptotic observers.

For example, based on the large volume compactification scenario \cite{Conlon:2006gv}, we adopt the following model:
\begin{eqnarray}
S_{\mathrm{LVC}} = \int dx^{4} \sqrt{-g} \left[ \frac{1}{16\pi} R - \frac{1}{2}\nabla_{\mu} \Phi \nabla^{\mu} \Phi - V(\Phi) + \beta e^{-c\Phi} \mathcal{L}_{\mathrm{M}} \right],
\end{eqnarray}
where $\Phi$ is the moduli field with the potential
\begin{eqnarray}
V=\left(1-\kappa \Phi^{3/2}\right)e^{-\sqrt{27/2}\Phi}+\epsilon e^{-\sqrt{6}\Phi}.
\end{eqnarray}
Now the matter sector $\mathcal{L}_{\mathrm{M}} = - \frac{1}{2} g^{\mu\nu} \phi_{;\mu} \phi_{;\nu} - \frac{1}{2} m^{2} \phi^{2}$ can interact with the moduli field and it can destabilize the moduli field, since the effective potential of the moduli sector is $V + \mathcal{L}_{\mathrm{M}} e^{-\sqrt{6} \Phi}$. Fig. \ref{fig:plots} is such an example \cite{Hwang:2013zaa}: a gravitational collapse can allow a locally destabilized moduli field, though eventually the moduli field will be stabilized after a long time. Since the destabilized region is outside the event horizon, it opens a possibility that a strong gravitational behavior near the event horizon can carry string theoretical effects.

\section{\label{sec:per}Perspectives}

In this review article, we summarized previous accomplishments of the double-null formalism. There is a wide range of applications, where one can extend the gravity sector as well as the matter sector. One can vary dimensions or symmetries and see various interesting applications. There are several topics that can only be understood by numerical simulations, e.g., cosmic censorship, mass inflation, and several semi-classical models regarding the information loss problem. These numerical data can be applied for the test of astrophysical experiments, e.g., gravitational wave physics, while the strong gravitational phenomena can include information of dark matter as well as string-inspired corrections.

Although there have been many applications, there are several directions that the double-null simulation should be updated.
\begin{itemize}
\item[--] Further extension to modified gravity: Up to now, modified gravity was limited to the Brans-Dicke type models. However, the formalism can be extended to higher derivation models, e.g., the Gauss-Bonnet-dilaton gravity or massive gravity.
\item[--] Further extension to modified matter: Like this, we have considered canonical form of matter fields. However, the formalism can be extended to non-canonical matter fields, for example, non-linear electrodynamics or $k$-essence model.
\item[--] Further extension to dimensions and topologies: What will be the phenomenology for higher dimensions? For the higher dimensions, one can give various topologies, for example some directions are non-compact and the other directions are compact. As one investigates strong gravitational phenomena with various topologies in higher dimensions, there can be various applications for the string theory.
\end{itemize}

As a theoretical black hole factory, the double-null formalism has been a very strong and useful tool. Also, it has a potential to remain a good frontier toward the complete understanding of quantum gravity, as it provides an effective way of investigating gravitational phenomena in the strong field regime. In addition, by using the detailed numerical calculations, we can find a connection with various observational consequences. In the end, we hope that the double-null formalism and numerical calculations contribute toward the ultimate understanding of the universe.

\section*{Acknowledgment}

AN was supported by the National Science Centre, Poland under a postdoctoral scholarship DEC-2016/20/S/ST2/00368. {\L}N was supported by the National Science Centre, Poland under a postdoctoral scholarship DEC-2017/26/D/ST2/00193. DY was supported by the Korean Ministry of Education, Science and Technology, Gyeongsangbuk-do and Pohang City for Independent Junior Research Groups at the Asia Pacific Center for Theoretical Physics and the National Research Foundation of Korea (Grant No.: 2018R1D1A1B07049126).

\end{document}